\documentstyle[aps,preprint]{revtex}
\begin{document}
\draft
\title{
Underpotential deposition of Cu on Au(111) in sulfate-containing
electrolytes: a theoretical and experimental study
}
\author{Jun Zhang,$^{1}$ Yung-Eun Sung,$^{2}$
Per Arne Rikvold,$^{1,3,4,*,\dag}$
and
Andrzej Wieckowski$^{2,\dag}$}
\address{
$^1$ Supercomputer Computations Research Institute,\\
Florida State University, Tallahassee, Florida 32306-4052\\
$^{2}$ Department of Chemistry
and Frederick Seitz Materials Research Laboratory,\\
University of Illinois, Urbana, Illinois 61801\\
$^{3}$ Center for Materials Research and Technology,
and Department of Physics,\\
Florida State University, Tallahassee, Florida 32306-3016\\
$^{4}$ Centre for the Physics of Materials and Department of Physics,\\
McGill University, Montr{\'{e}}al, Qu{\'{e}}bec, CANADA$^{\ddag}$
}
\date{\today}
\maketitle
\begin{abstract}
We study the underpotential deposition of Cu on single-crystal Au(111)
electrodes in sulfate-containing electrolytes by a combination of
computational
statistical-mechanics based lattice-gas modeling and experiments.
The experimental methods are {\it in situ\/} cyclic voltammetry and coulometry
and {\it ex situ\/} Auger electron spectroscopy and low-energy electron
diffraction. The experimentally obtained voltammetric current and charge
densities and adsorbate coverages are compared with the predictions of a
two-component lattice-gas model for the coadsorption of Cu and sulfate.
This model
includes effective, lateral interactions out to fourth-nearest neighbors.
Using group-theoretical ground-state calculations and Monte Carlo simulations,
we estimate effective electrovalences and lateral adsorbate--adsorbate
interactions so as to obtain overall agreement with experiments,
including both our own and those of other groups. In agreement with earlier
work, we find a mixed ($\sqrt3$$\times$$\sqrt3$)
phase consisting of 2/3 monolayer Cu and 1/3 monolayer sulfate at intermediate
electrode potentials, delimited by phase transitions at both higher and lower
potentials. Our approach provides estimates of the effective
electrovalences and lateral interaction energies,
which cannot yet be calculated by first-principles methods.
\end{abstract}


\section{Introduction}
\label{secint}

Underpotential deposition (UPD) is the process by which a submonolayer or
monolayer of one metal is electrochemically adsorbed onto another
at electrode potentials that are positive with respect to the potential
for bulk deposition. It occurs when the adsorbate adatoms are
more strongly bound to the foreign substrate than to a substrate of their
own kind \cite{KOLB78}. This phenomenon has been known since the early
years of this century \cite{HEVE12,ROGE49,MILL53,DEGE59},
but its study has intensified only during the last two decades
\cite{KOLB78,KOLB91}.
Current-potential data for UPD on single-crystal surfaces,
obtained from cyclic voltammetry (CV) experiments, frequently display sharp
current peaks \cite{SCHU76,DICK76}.
The shapes, positions, and number of these peaks depend on the substrate and
the crystal plane on which the adsorption takes place
\cite{SCHU76,DICK76,MAGN91},
as well as on the nature of the electrolyte \cite{ZEI87}.
It was already suggested in Refs.~\cite{SCHU76,DICK76,BECK77}
that the CV current
peaks separate ordered and disordered adsorbate structures,
and it was later pointed out that the peaks themselves should
correspond to phase transitions in the adsorbed layer
\cite{KOLB78,HUCK91,BLUM91}.
The UPD phenomenon is of fundamental interest as an aspect of the detailed,
microscopic structure of electrode--electrolyte interfaces, which is presently
the focus of vigorous research \cite{TRAS95}. It provides a means of
controlling the microscopic
surface structure through the electrolyte composition and
the applied potential, and
likely applications are suggested by the observations
that metallic submonolayers have the ability to significantly modify
the catalytic activity of a substrate \cite{ADZI84},
and that UPD
is the first step in the liquid-phase epitaxial growth of metallic
heterostructures \cite{DHEZ88}.

The phase-transition aspects of UPD make this area of surface electrochemistry
well suited for statistical-mechanical modeling
\cite{HUCK90,BLUM90,RIKV95,RIKV96}.
The first detailed, statistical-mechanical lattice-gas model for a UPD
system was introduced by Huckaby and Blum (HB),
who considered the electrosorption
of copper (Cu) onto the (111) plane of gold [Au(111)] in the presence
of sulfate
\cite{HUCK91,BLUM91,HUCK92,BLUM93,BLUM94A,BLUM94B,HUCK95,BLUM96}.
This system also has been intensively studied by a number of experimental
techniques (see Sec.~\ref{sex}), and the detailed data that are
now available render it an excellent candidate for theoretical modeling.

In this article we present a study of the UPD of Cu on well-characterized,
single-crystal Au(111) electrodes in sulfate-containing electrolytes,
in which we employ both experimental and theoretical methods.
The experimental techniques used are {\it in situ\/} CV and coulometry
to detect the current peaks and measure the total charge transferred
during the adsorption/desorption processes,
and {\it ex situ\/} Auger electron spectroscopy (AES) and low-energy electron
diffraction (LEED) to determine the adsorbed Cu and sulfate coverages
and the adlayer structure, respectively.
In the theoretical part of the study we formulate a
full, two-component lattice-gas model with effective, lateral
interactions out to
fourth-nearest neighbors. This is a generalization of the HB model
\cite{HUCK91,BLUM91,HUCK92,BLUM93,BLUM94A,BLUM94B,HUCK95,BLUM96}.
Using group-theoretical ground-state
calculations at zero temperature \cite{DOMA78,DOMA79,SCHI81}
and Monte Carlo (MC) simulations at room temperature
\cite{BIND86,BIND92,BIND92B},
we estimate the model parameters (the effective, lateral
interactions and electrovalences) so as to obtain overall agreement
with the available experimental data, including both our own and those of
other groups. This theoretical approach has
previously been applied to electrochemical problems by Rikvold and co-workers
\cite{RIKV88A,RIKV88B,COLL89,RIKV91A,RIKV91B},
and it was most recently used to model the coadsorption of hydrogen
with urea on single-crystal platinum (100) electrodes
\cite{RIKV95,RIKV92,RIKV93A,RIKV93B,GAMB93B,HIGH93,GAMB94}
and with sulfate on rhodium (111) \cite{RIKV95}.
It differs from the approach taken by HB
in that we apply the full, two-component lattice-gas model to
analyse {\it global\/}, as well as local features of the system's phase
diagram and CV current profiles,
and in that our room-temperature results are obtained by nonperturbative,
numerical simulation.
(For reviews and discussion of the HB approach, see
Refs.~\cite{HUCK95,BLUM96}.)
Some preliminary results of this study have been reported in
Refs.~\cite{RIKV96,JZHA95}.

The remainder of this paper is organized as follows.
In Sec.~\ref{sex} we give a survey of experimental results
that are currently available.
In Sec.~\ref{secexp} we describe our experimental procedures and results.
In Sec.~\ref{secmod} we introduce our lattice-gas model and give the relations
between the lattice-gas parameters and the experimentally observable
quantities. Here we also discuss our procedure for adjusting the model
parameters to fit the experiments.
In Sec.~\ref{secres} we present together our numerical and experimental
results, and in
Sec.~\ref{secFSS} we give a finite-size scaling analysis of the
phase transitions predicted by our model.
In Sec.~\ref{secD} we briefly summarize our conclusions
and discuss their significance.

\section{Survey of the Experimental Situation}
\label{sex}

The UPD of Cu on Au(111) in sulfate-containing electrolytes has been
extensively studied, both by {\it in situ\/} techniques like CV
\cite{SCHU76,DICK76,ZEI87,BECK77,HACH91,OMAR93,BORG94%
,ZSHI94B,ZSHI94C,ZSHI95,ZSHI95B},
chronocoulometry
\cite{HACH91,OMAR93,BORG94,ZSHI94B,ZSHI94C,ZSHI95,ZSHI95B},
scanning tunneling microscopy (STM) \cite{MAGN91,HACH91,MAGN90,KOLB94},
atomic-force microscopy (AFM) \cite{MANN91},
Fourier-transform infrared spectroscopy (FTIR) \cite{PARR93},
X-ray-absorption spectroscopy (EXAFS and XANES)
\cite{BLUM86,MELR88,TADJ91},
surface X-ray scattering \cite{GORD94,TONE95},
quartz crystal microbalance (QCM) \cite{BORG94,GORD94},
and by {\it ex situ\/} techniques,
such as LEED, RHEED (reflection high-energy electron diffraction) and AES
\cite{ZEI87,BECK77,NAKA84,KOLB86,KOLB87}.
Electrolyte compositions used in several CV experiments (including the
present work) are
listed in Table~\ref{table1}, together with potential scan rates and
observed current-peak separations. The observation
that the different electrolyte
compositions included in Table~\ref{table1} do not qualitatively change the
shape of the voltammogram indicates that within this range of compositions the
adsorbates remain the same. In this paper we assume these to be
Cu$^{2+}$ and SO$_4^{2-}$ in solution, which are adsorbed as near-neutral
Cu and partially discharged or neutralized SO$_4^{2-}$, respectively
\cite{ZSHI94B,ZSHI94C,ZSHI95,ZSHI95B,ZSHI94D}.

A three-step experiment was performed by Shi and Lipkowski
\cite{ZSHI94B,ZSHI94C} to clarify the roles of
the different adsorbates. With 0.1 M perchloric acid (HClO$_4$)
as the only electrolyte ({\it i.e.}, with no copper added), the CV current
in the double-layer range was weak, indicating that the ions present in
the solution did not adsorb significantly on the surface. In the next step,
K$_2$SO$_4$ was added, introducing HSO$_4^-$ and SO$_4^{2-}$ ions into the
solution. An adsorption peak then appeared, particularly
during the positive-going scan. Radiochemical studies \cite{ZSHI94D} indicate
that this peak is due to sulfate adsorption with a maximum coverage of
approximately 0.2 monolayers (ML).
This result agrees with chronocoulometric results \cite{ZSHI94D}
and with earlier STM studies
\cite{MAGN92,EDEN94},
in which a
$(\sqrt{3}\!\times\!\sqrt{7})$
sulfate adlayer was indicated on the
positive side of the voltammetric peak in the absence of copper,
without \cite{MAGN92} or with \cite{EDEN94}
a perchloric-acid supporting electrolyte.
(In UHV studies with samples emersed  at potentials for which the
$(\sqrt{3}\!\times\!\sqrt{7})$
structure is observed {\it in situ\/}, only a
$(\sqrt{3}\!\times\!\sqrt{3})$
surface structure is observed, presumably produced by desorption of
water that was coadsorbed with the sulfate \cite{MROZ94}.)
Finally, addition of Cu$^{2+}$ [as Cu(ClO$_4$)$_2$]
caused the appearance of two sharp (10--30~mV wide) current
peaks that totally dominate the voltammogram, as illustrated
by our own data in Fig.~\ref{eclr}(a).
There is no two-peak feature in the CV if no copper is added.
On the other hand, if the copper concentration is kept fixed
while the concentration of sulfate is reduced, either the peak heights are
reduced, or the two peaks merge, or both effects are observed together
\cite{ZSHI94B,ZSHI94C}. Evidently, the coadsorption of Cu and sulfate
is mutually enhancing.
(The same effects, albeit with a smaller peak separation, are also
seen with
Pt(111) electrodes
in the same electrolytes
\cite{SHIN95}.)

In Fig.~\ref{eclr}(b) is shown the charge density transferred during the
adsorption/desorption process, obtained as the integral of the CV current
density in Fig.~\ref{eclr}(a) between +120 and +420~mV {\it vs.\/} Ag/AgCl.
The charge plateau in the potential region between the CV peaks
(approximately 200--300~mV)
lies about 200~$\mu$C/cm$^2$
below the charge density at 420~mV.
If the reactions Cu $-$ 2e$^-\!\rightarrow$ Cu$^{2+}$ and
SO$_4$ + 2e$^-\!\rightarrow$ SO$_4^{2-}$
are assumed across the right-hand peak (Peak~\#1),
the desorption of 2/3~ML of Cu and 1/3~ML of
sulfate would result in a total charge transfer in that potential region of
160~$\mu$C/cm$^2$ \cite{OMAR93}.
This coverage--charge correspondence is supported by
RHEED and LEED
\cite{BECK77,NAKA84}, by which a ($\sqrt{3}\!\times\!\sqrt{3}$)
superstructure was observed in the potential region between the peaks.
Further, consistent results have been obtained by STM
\cite{MAGN91,HACH91,MAGN90,KOLB94}.
Corroborating evidence was also obtained in an {\it in situ\/} electrochemical
QCM study \cite{BORG94}.
At the left-hand peak (Peak~\#2) and for still
more negative potentials, the Cu coverage approaches 1~ML, and
EXAFS results have suggested that sulfate adsorbs on top of the
Cu atoms when a monolayer of Cu has been deposited on the Au(111) surface
\cite{BLUM86}.
This interpretation is
supported by the measurements of Shi and Lipkowski \cite{ZSHI94C}
and by our own AES results.
The total change can therefore not be explained
by the specified charge-transfer
reactions, which would predict a much larger charge
transfer than the one measured.
Shi and Lipkowski \cite{ZSHI94B,ZSHI94C}
derived an equilibrium electrocapillary equation
and determined the relative interfacial tension from measurements
of the electrode charge density, using the back-integration technique.
The Cu coverage is then simply obtained as the derivative of the
relative interfacial tension with respect to the electrochemical potential of
Cu. They compared the corresponding
charge with the calculated Cu coverage
and concluded that the effective electrovalence of copper varies from
1.6 to 1.8 in the double-peak range of the electrode potential.
These results are consistent with those obtained in our present work.

The total charge transfer measured from CV results varies among the reported
experiments. In the potential range from
120 to 420~mV {\it vs.\/} Ag/AgCl, which includes both
CV peaks, some of the available coulometric results give values
about $-$350~$\mu$C/cm$^2$ \cite{OMAR93,ZSHI94B,ZSHI94C,ZSHI95,ZSHI95B}
(also the result of the present work),
$-$460~$\mu$C/cm$^2$ \cite{HACH91},
and $-$470~$\mu$C/cm$^2$ \cite{BORG94},
compared with the theoretical value of $-$444~$\mu$C/cm$^2$
for a monolayer of ions transferring two electrons per ion to the
Au(111) surface. According to Refs.~\cite{ZSHI94B,ZSHI94C},
variation of the sulfate concentration
from 0.01~mM to 5~mM changes the total charge transfer
only by about 5\% at the negative end of the UPD potential range.
Likewise, variation of the Cu(ClO$_4)_2$ concentration from 0.01~mM to 5~mM
also did not yield significant changes in the total charge.
Thus, it is highly unlikely that
the large discrepancies between the experimentally measured charge transfers
are due to the differences between the electrolyte compositions used.
Rather, we believe
that the large charge transfers observed in Refs.~\cite{HACH91,BORG94}
were due to unexpectedly large roughness of the electrodes.

The only systematic studies of kinetic aspects of the UPD in this system of
which we are aware, are by H{\"o}lzle, {\it et al.\/}
\cite{KOLB94,HOLZ94A,HOLZ94}. They considered the dependence on the potential
scan rate of the the separation between
the positions of the positive- and negative-going CV current peaks
({\it i.e.\/}, the hysteresis) and the corresponding maximum current
densities. In both cases they found a linear dependence on the square root
of the scan rate. Additional potential-step measurements, in
which the potential was changed discontinuously from one side of a peak to
the other, gave indications of first-order phase transitions and
phase change via a heterogeneous nucleation mechanism.

\section{Experimental Procedures and Results}
\label{secexp}

In this section we detail the experimental procedures employed in the present
study, and we present our experimental results. In later sections
these will be used to formulate our lattice-gas model and determine
its parameters.

The combined ultra-high vacuum (UHV) and electrochemistry instrument
has been described elsewhere \cite{MROZ94,KAMR90,SUNG94}.
Prior to the electrochemical experiments, the (111)
plane of the gold single crystal (Johnson Matthey, 99.995\% purity) was
cleaned by 340~eV Ar$^+$ ion sputtering and annealed at around 650~K.
The meniscus position of the working electrode ensured that only the chosen
plane of the gold crystal was exposed to the electrolyte.  Following removal
of the electrode from solution, a drop of electrolyte remained hanging from
the surface.  The drop was removed using Teflon tubing terminated with a
syringe \cite{SUNG94}.
The electrode was next transferred to UHV for characterization
by electron spectroscopies.  There was no carbon contamination on the
emersed surfaces, and gold surfaces showed complete water desorption
\cite{THIE88}.

The electrochemical measurements were conducted using conventional
three-electrode circuitry and an EG\&G PAR 362 potentiostat. The working
solutions were made of Millipore water (18 M$\Omega \cdot$cm),
ultra-pure grade H$_2$SO$_4$
(Ultrex from VWR), and ultra-pure grade HClO$_4$ (VWR).  Solutions were
deaerated and blanketed with nitrogen (Linde, Oxygen Free, 99.99\%).
The CuSO$_4$ (99.999\%, Aldrich)
and Na$_2$SO$_4$ (ACS grade, Fisher Scientific) chemicals
were used as received.
Voltammetric current profiles obtained at a scan rate of 2~mV/s are shown in
Fig.~\ref{eclr}.
The separations between the positive- and negative-going CV peak potentials
were 24~mV for Peak~\#1 and 44~mV for Peak~\#2, indicating that kinetic
effects \cite{KOLB94,HOLZ94A,HOLZ94}
may contribute to the shape of the voltammogram to a
nonnegligible degree, even at this relatively low scan rate.

The AES was carried out in differentiated mode
with 2~eV modulation amplitude, using a Perkin Elmer PHI-10-155 cylindrical
mirror analyzer (CMA).  The spectra were obtained using a digital data
acquisition system, and smoothed one time using a simple 11-point averaging
technique.  The analysis was performed within 5--8 seconds at a given point
on the sample, after which the location on the sample was changed for a
subsequent 5--8 seconds' measurement (to reduce the electron-beam damage).
These AES spectra are shown in Fig.~\ref{figA2}.

Quantitative AES analyses of sulfate were performed using a standardization
technique developed by Wieckowski and co-workers \cite{MROZ94,KAMR90,SUNG94}.
In this procedure, a thick
deposit of sodium sulfate was formed after water evaporation from gold
emersed from aqueous 2~M Na$_2$SO$_4$.
The Au(111) surface covered by Na$_2$SO$_4$ was
transferred into UHV and used as a standard for the AES characterization of
the electrode covered by sulfate up to 1~ML coverage. The thick,
bulk-like layer of Cu deposited from a solution
containing 2~M Cu$^{2+}$ and 1~mM H$_2$SO$_4$
at a negative potential served as the reference standard for the Cu
intensity. We used a sensitivity factor approach within a homogenous model
of the adsorbed layer, resulting from nonlinear equations discussed
in Ref.~\cite{MROZ94}.
The resulting coverages of sulfate
[obtained from the S(LMM) Auger electron transitions] and
Cu [from the Cu(LMM) and Cu(MNN) transitions]
are shown {\it vs.\/}
the electrode potential, together with our simulated coverage
results, in Fig.~\ref{cg}.

Our measurements were conducted in a very diluted sulfuric acid solution
(0.1~mM), containing 1~mM of copper sulfate.  The low bulk concentration of
copper cations ensures that practically none of the Cu-species remains in the
emersed film, that could be added to the adsorbate coverage measured by AES
in UHV.  Therefore,
no rinsing of the electrode surface before emersion to vacuum was
necessary. The spot width of the hexagonal (111) LEED pattern of
the surface indicates a (1$\times$23) reconstruction of the Au(111) surface,
which is confirmed by the sharp CV spike in pure H$_2$SO$_4$ solution,
seen in Fig.~\ref{eclr}(a)
at approximately 440~mV {\it vs.\/} Ag/AgCl \cite{MROZ94,MROZ95}.
A representative LEED pattern is shown in Fig.~\ref{figA1}.

All measurements were conducted at room temperature.  Unless otherwise
indicated, the electrode potentials are referred to Ag/AgCl.

\section{Model and Theoretical Methods}
\label{secmod}

Our lattice-gas model for the UPD of Cu on Au(111) in sulfate-containing
electrolyte is a
refinement of the HB model
\cite{HUCK91,BLUM91,HUCK92,BLUM93,BLUM94A,BLUM94B,HUCK95,BLUM96}.
It is based on the assumption that the
sulfate
coordinates the triangular
Au(111) surface
(STM indicates that the (1$\times$23) reconstruction observed by LEED
(see Sec.~\ref{secexp}) does not occur {\it in situ\/} \cite{MAGN91})
through three of its
oxygen atoms, with the fourth S-O bond pointing away from the surface
\cite{HUCK91,HUCK92,BLUM94A,BLUM94B,TONE95,EDEN94} (as is
also the most likely adsorption geometry on Rh(111) \cite{RIKV95}).
This adsorption
geometry gives the sulfate a ``footprint" in the shape of an
equilateral triangle with a O-O distance of 2.4 \AA \,
\cite{PASC65},
reasonably matching
the lattice constant for the triangular Au(111) unit cell, 2.88 \AA \,
\cite{KITT86}.
The Cu is assumed to compete for the same adsorption sites as the sulfate.
The configuration energies of the coadsorbed particles are given by
the standard three-state
lattice-gas Hamiltonian (energy function) used, {\it e.g.}, in
Refs.~\cite{RIKV88A,RIKV88B,COLL89,RIKV91A,RIKV91B}:
\begin{eqnarray}
{\cal H}_{\rm LG}
&=& \sum_l \Bigg[ -\Phi_{\rm SS}^{(l)}
        \sum_{\langle ij \rangle}^{(l)} n_i^{\rm S} n_j^{\rm S}
       -\Phi_{\rm SC}^{(l)} \sum_{\langle ij \rangle}^{(l)}
        \left(n_i^{\rm S} n_j^{\rm C}
       + n_i^{\rm C} n_j^{\rm S} \right)
       -\Phi_{\rm CC}^{(l)}
        \sum_{\langle ij \rangle}^{(l)} n_i^{\rm C} n_j^{\rm C} \Bigg]
\nonumber\\
& & - \Phi_{\rm SS}^{(t)}\sum_{\triangle}n_in_jn_k
    - \bar{\mu}_{\rm S} \sum_i n_i^{\rm S}
    - \bar{\mu}_{\rm C} \sum_i n_i^{\rm C}  \; .
\label{eq1}
\end{eqnarray}
Here $n_i^{\rm X}\in$\{0,1\}
is the local occupation variable for species X
[X=S (sulfate) or~C (copper)], and the third adsorption state
(``empty'' or ``solvated'')
corresponds to $n_i^{\rm S}$=$n_i^{\rm C}$=0.
The sums $\sum_{\langle ij \rangle}^{(l)}$ and $\sum_i$ run over all
$l$th-neighbor bonds and over all adsorption sites,
respectively, $\Phi_{\rm XY}^{(l)}$ denotes the
effective XY pair interaction corresponding to an
$l$th-neighbor bond, and $\sum_l$ runs over the interaction ranges.
The term $-\Phi_{\rm SS}^{(t)}\sum_\triangle n_in_jn_k$ expresses
three-particle interactions between sulfates, involving all
second-neighbor equilateral triangles
\cite{EINS79,PAYN92,HUCK92B}.
Those bonds that correspond to potentially
nonzero lateral interactions in the present implementation
of our model are shown in
Fig.~\ref{pic1}.
For large separations the interactions are assumed to vanish,
and $\Phi_{\rm SS}^{(1)}$ is an infinite repulsion corresponding
to nearest-neighbor sulfate-sulfate exclusion
(``hard hexagons"
\cite{HUCK92,BLUM94A,BLUM94B,BAXT82}).
The change in electrochemical potential when one X
particle is removed from the bulk solution and adsorbed on the surface
is $-\bar{\mu}_{\rm X}$.
The sign convention is such that $\Phi_{\rm XY}^{(l)}$$>$0
denotes an effective attraction, and $\bar{\mu}_{\rm X}$$>$0
denotes a tendency for adsorption in the absence of lateral
interactions. We emphasize that the $\Phi_{\rm XY}^{(l)}$ are {\it effective}
interactions through several channels, including electron-, phonon-,
electrostatic, and fluid-mediated mechanisms
(see Refs.~\cite{RIKV96,RIKV91A} for discussion and references).

The electrochemical potentials in Eq.~(\ref{eq1}) are (in the weak-solution
approximation and here given in molar units) related to the bulk
concentrations [X] and the electrode potential $E$ as
\begin{eqnarray}
\bar{\mu}_{\rm S} & =
         & \mu_{\rm S}^0+RT\ln \frac{[{\rm S}]}{[{\rm S}]^0}-z_{\rm S}FE
\nonumber \\
\bar{\mu}_{\rm C} & =
         & \mu_{\rm C}^0+RT\ln \frac{[{\rm C}]}{[{\rm C}]^0}-z_{\rm C}FE\,.
\label{chempo}
\end{eqnarray}
Here $z_{\rm S}$ and $z_{\rm C}$ are the effective electrovalences of
sulfate and copper, respectively,
$R$ is the molar gas constant, $T$ is the
absolute temperature, and $F$ is the Faraday constant. The quantities
superscripted with a 0 are reference values which contain the local binding
energies to the surface. They are generally temperature dependent due,
among other effects, to rotational and vibrational modes.

The coverages of sulfate and Cu, $\Theta_{\rm S}$ and $\Theta_{\rm C}$,
are the thermodynamic densities conjugate to $\bar{\mu}_{\rm S}$ and
$\bar{\mu}_{\rm C}$, respectively. They are defined as
\begin{equation}
\Theta_{\rm X} = N^{-1}\sum n_i^{\rm X},
\end{equation}
where $N$ is the total number of unit cells in the surface lattice.
However, it is experimentally observed that the sulfate species
remains partially adsorbed on top of the complete monolayer of Cu
in the negative-potential region, rather than becoming desorbed
into the solution \cite{ZSHI94C,BLUM86},
as was mentioned in Sec.~\ref{sex}.
Rather than introducing a full, two-layer lattice-gas model, we
approximate the sulfate coverage in this second layer as
\begin{equation}
\label{eqTh2}
\Theta_{\rm S}^{(2)} = \alpha\Theta_{\rm C}(1/3-\Theta_{\rm S}) \;,
\end{equation}
which allows the difference between the first-layer coverage $\Theta_{\rm S}$
and its saturation value of 1/3 to be transferred to the top of the Cu
layer. The factor $\alpha$ is a phenomenological fitting parameter expected
to be between zero and one. The total sulfate coverage measured, {\it e.g.},
in chronocoulometric \cite{ZSHI94C} and AES experiments, is
$\Theta_{\rm S}+\Theta_{\rm S}^{(2)}$, and the total charge transported
across the surface per unit cell during the adsorption/desorption process
thus becomes
\begin{eqnarray}
\label{eqQ}
q & = & -e\left[ z_{\rm C}\Theta_{\rm C}
        + z_{\rm S}\left(\Theta_{\rm S}+\Theta_{\rm S}^{(2)}\right)\right]
                                                              \nonumber \\
  & = & -e\left[(z_{\rm C}+\alpha z_{\rm S}/3)\Theta_{\rm C}
-z_{\rm S}\alpha\Theta_{\rm C}\Theta_{\rm S}
+ z_{\rm S}\Theta_{\rm S})\right],
\end{eqnarray}
 where $e$ is the elementary charge unit.
In the absence of diffusion and double-layer effects and in the limit that
the potential scan rate ${\rm d}E/{\rm d}t\!
\rightarrow\! 0$
\cite{BARD80},
the voltametric
current $i$ per unit cell of the surface is the time derivative of $q$.
Using differentiation by parts involving the relations between the electrode
potential and the electrochemical potentials, Eq.~(\ref{chempo}), as well
as the Maxwell relation
$\partial \Theta_{\rm S}/\partial \bar{\mu}_{\rm C} =
\partial \Theta_{\rm C}/\partial \bar{\mu}_{\rm S}$,
we find the current density $i$ in terms of the lattice-gas
response functions $\partial \Theta_{\rm X} / \partial \bar{\mu}_{\rm Y}$:
\begin{eqnarray}
i & = & eF\left\{ z_{\rm S}^2(1-\alpha\Theta_{\rm C})
\left.\frac{\partial\Theta_{\rm S}}
{\partial\bar{\mu}_{\rm S}}\right|_{\bar{\mu}_{\rm C}}
+z_{\rm C}(z_{\rm C}-2\alpha z_{\rm S}\Theta_{\rm S}/3)
\left.\frac{\partial \Theta_{\rm C}}
{\partial \bar{\mu}_{\rm C}} \right|_{\bar{\mu}_{\rm S}} \right.  \nonumber \\
  &   & +
\left.
z_{\rm S}(2z_{\rm C}+\alpha z_{\rm S}(1/3-\Theta_{\rm S})-\alpha z_{\rm C}
\Theta_{\rm C})
\left.\frac{\partial\Theta_{\rm S}}
{\partial \bar{\mu}_{\rm C}}\right|_{\bar{\mu}_{\rm S}}
\right\}\frac{{\rm d}E}{{\rm d}t}\,,
\label{fd}
\end{eqnarray}
which reduces to its standard form for $\alpha=0$
\cite{RIKV95,RIKV93A,RIKV93B,GAMB93B}.
We emphasize that the derivation of Eq.~(\ref{fd}) is based on
{\it equilibrium\/}
statistical mechanics. Very close to a phase transition it
may require exceedingly slow scan rates to be valid, due to critical
slowing-down in the case of second-order transitions
\cite{BIND86,BIND92,BIND92B}
or hysteresis in the case of  first-order transitions
\cite{KOLB94,HOLZ94A,HOLZ94,RIKV94}.

The following steps are involved in adjusting the lattice-gas parameters
to fit the experimental data.\\
{\bf A.}
Ground-state calculations \cite{DOMA78,DOMA79,SCHI81}
to determine parameter ranges that yield
ground states which correspond to the experimentally observed phases
at room temperature \cite{HUCK95}.\\
{\bf B.}
MC simulations \cite{BIND86,BIND92,BIND92B} at room temperature
to further adjust the effective interactions to obtain reasonable
agreement with the shapes of the
observed adsorption isotherms and CV profiles.\\
{\bf C.}
With the effective interactions obtained in step {\bf B}, determination of the
effective electrovalences so as to give agreement between the dependences
of the CV peak positions on the electrolyte composition in
experiments and simulations.\\
{\bf D.}
Iteration of steps {\bf B} and {\bf C}
as necessary to improve the agreement between
the experiments and simulations.

\subsection{Ground-State Calculations}
\label{secgs}

Ordered phases with unit cells up to ($3 \! \times \! 3$)
were identified by applying the group-theoretical
arguments of Landau and Lifshitz
\cite{DOMA78,DOMA79,SCHI81}.
The ground-state energy of each phase
depends on $\bar{\mu}_{\rm S}$ and $\bar{\mu}_{\rm C}$ and the lateral
interactions.
Among phases with identical coverages of Cu
and sulfate, those were excluded that have a higher ground-state energy
than others with the same coverages for all chemically reasonable choices
of interaction energies.
Out of the 44 remaining phases, ten were found to be
realized as ground states for interactions in the
range of experimental interest. These are listed in Table~\ref{t2},
where they are identified by the notation
($X\!\times\!Y$)$_{\Theta_{\rm C}}^{\Theta_{\rm S}}$.
For a given set of
interactions, the zero-temperature phase boundaries were exactly determined
by pairwise equating the energies given in Table~\ref{t2}. In order to easily
explore the effects of changing one or more of the interactions, a
computer program was
developed which numerically determines the ground-state diagram by
scanning $\bar{\mu}_{\rm S}$ and $\bar{\mu}_{\rm C}$ and determining the
phase of minimum ground-state energy
\cite{BUEN94}.

The ground-state diagram obtained with the final set of
interaction constants used in this work is shown in Fig.~\ref{gs}.
For large negative $\bar{\mu}_{\rm S}$ only Cu adsorption is possible,
and the phase diagram is that of the lattice-gas model corresponding to the
triangular-lattice antiferromagnet with next-nearest neighbor ferromagnetic
interactions
\cite{LAND83,KITA88,DEQU95}.
Similarly, in the limit of large positive $\bar{\mu}_{\rm S}$
and large negative $\bar{\mu}_{\rm C}$ the zero-temperature phase is the
$(\sqrt{3}\!\times\!\sqrt{3})_0^{1/3}$ sulfate phase characteristic
of the hard-hexagon
model \cite{HUCK91,HUCK92,BLUM94A,BLUM94B,BAXT82}.
The phase diagram for intermediate electrochemical
potentials is quite complicated.
For $\bar{\mu}_{\rm S} < -22$ kJ/mol, no sulfate adsorption occurs in the
first adlayer,
while for $\bar{\mu}_{\rm C}\! <\! -18$ kJ/mol, no Cu is adsorbed.
The $(\sqrt{3}\!\times\!\sqrt{3})_{2/3}^{1/3}$
mixed-phase region in the upper right-hand part of the diagram is
relatively large,
due to the nearest-neighbor effective attraction between Cu and sulfate,
$\Phi_{\rm SC}^{(1)}$. This is the mixed phase which occurs between the
CV peaks.
The $(\sqrt{3}\!\times\!\sqrt{7})_0^{1/5}$ phase corresponds to experimental
observations in copper-free systems \cite{MAGN90,EDEN94}.
This phase is enhanced by
the fourth-neighbor sulfate-sulfate attraction, $\Phi_{\rm SS}^{(4)}$, and
the $(\sqrt{3}\!\times\!\sqrt{3})_0^{1/3}$ (``hard-hexagon'')
phase (which has not been directly observed in this system) is disfavored
by $\Phi_{\rm SS}^{(t)}$, the second-neighbor repulsive trio interactions.
The $(\sqrt{3}\!\times\!\sqrt{7})_0^{1/5}$ phase is also very sensitive to
the next-nearest neighbor interaction $\Phi_{\rm SS}^{(2)}$.
Attractive values of this interaction favor the
$(\sqrt{3}\!\times\!\sqrt{3})_0^{1/3}$ phase,
whereas repulsive values favor the
$(\sqrt{7}\!\times\!\sqrt{7})_0^{1/7}$ phase.
The experimental observations of the
$(\sqrt{3}\!\times\!\sqrt{7})_0^{1/5}$ phase therefore indicate that
$\Phi_{\rm SS}^{(2)}$ must be much weaker than the other effective
interactions, and we set it equal to zero in our simulations.

\subsection{Simulations at Room Temperature}
\label{secMC}

At nonzero temperatures, the thermodynamic and structural properties of
a lattice-gas model can be obtained by a number of
analytical and numerical methods. These include mean-field
approximations \cite{HOMM89,ARMA91A},
Pad{\'e}-approximant methods based on liquid theory
\cite{HUCK91,BLUM91,HUCK90,HUCK92,BLUM93,BLUM94A,BLUM94B},
numerical transfer-matrix calculations
\cite{RIKV88A,RIKV88B,COLL89,RIKV91A,RIKV91B,RIKV92},
and MC simulations
\cite{RIKV95,RIKV96,COLL89,RIKV93A,RIKV93B,GAMB93B,HIGH93,GAMB94,JZHA95}.
The reason for our choice of the numerical MC method
\cite{BIND86,BIND92,BIND92B}
is that such non-perturbative methods as MC and transfer-matrix
calculations provide much more accurate results
for two-dimensional systems than even
quite sophisticated mean-field approximations \cite{RIKV93D}.
Yet they are quite easy to program, and with modern computer technology their
implementation is well within even modest computational resources.
(All our numerical data were produced on a cluster of
IBM RS/6000 workstations.)
For models with such relatively long-ranged interactions as the present one,
MC is better suited than numerical transfer-matrix methods.

The ground-state diagram serves as a guide to the path
in the ($\bar{\mu}_{\rm S}$, $\bar{\mu}_{\rm C}$) plane that the
room-temperature isotherms should follow. At constant temperature and {\it p}H,
two factors influence the path: the adsorbate concentrations in the bulk,
and the effective electrovalences.
As seen from Eq.~(\ref{chempo}), $\bar{\mu}_{\rm S}$ and
$\bar{\mu}_{\rm C}$ depend linearly on $E$, with proportionality constants
$-$$z_{\rm S}$ and $-$$z_{\rm C}$, which must be determined from experiments.
In order to obtain the parts of the room-temperature phase diagram of
the model that are
relevant to the experiments, we performed initial isothermal potential
scans, using the approximate values $z_{\rm S}=-2$
and $z_{\rm C} =+2$, which
correspond to the full ionic charges of SO$_4^{2-}$
and Cu$^{2+}$, respectively.
The experimental evidence, in particular Refs.~\cite{ZSHI94B,ZSHI94C,ZSHI95B},
indicates that these `ideal' values are somewhat too large.
After obtaining the room-temperature phase boundaries
as discussed in this Subsection, we therefore re-adjusted our estimates of
$z_{\rm S}$ and $z_{\rm C}$ as discussed in Subsec.~\ref{secZZ} below.

The MC simulations used
to obtain adsorption isotherms and CV currents at room temperature
($RT = 2.50$kJ/mol, corresponding to 28$^\circ$C)
were performed on triangular lattices of size
$L \! \times \! L$, using a heat-bath algorithm
\cite{BIND86,BIND92,BIND92B}
with updates at randomly chosen sites.
The majority of our simulations were performed with $L$=30,
supplemented with simulations for $L$=45 and~15.
Simulation results reported without specifying the system size
are for $L$=30.

In order to avoid getting stuck in
metastable configurations (a problem which is exacerbated by the
nearest-neighbor sulfate-sulfate exclusion),
we simultaneously updated clusters
consisting of two nearest-neighbor sites.
Each data point was obtained from a run of $10^5$ Monte Carlo
steps per site (MCSS),
starting from the appropriate ground-state configuration.
Sampling was performed at intervals of 50~MCSS,
and the first 4000~MCSS were discarded to ensure equilibration.
The coverages were obtained as averages over these samples, and the
nonordering lattice-gas response functions,
$\partial \Theta_{\rm S}/\partial \bar{\mu}_{\rm S}$,
$\partial \Theta_{\rm C}/\partial \bar{\mu}_{\rm C}$, and
$\partial \Theta_{\rm S}/\partial \bar{\mu}_{\rm C} =
\partial \Theta_{\rm C}/\partial \bar{\mu}_{\rm S}$,
and the adsorbate heat capacity
were calculated from the equilibrium coverage and energy
fluctuations in the standard way \cite{BIND86,BIND92,BIND92B}.

The lattice-gas interaction energies were varied until the widths of the
CV peaks corresponding to the room-temperature phase boundaries were
in acceptable agreement with the experimentally observed peak widths.
The corresponding peak positions are superimposed on
the ground-state diagram in Fig.~\ref{gs}.

\subsection{Effective Electrovalences}
\label{secZZ}

Once the room-temperature phase boundaries for the chosen set of interaction
constants had been calculated,
the effective electrovalences could be re-estimated by comparing
the simulated phase boundaries (CV peak positions) in the
$(\bar{\mu}_{\rm S},\bar{\mu}_{\rm C})$ plane with the dependence
on the electrolyte composition of experimentally observed CV peak potentials.
For this purpose we used data from Omar {\it et al.\/} \cite{OMAR93},
who measured the CV peak positions {\it vs.\/} the bulk concentration of
Cu$^{2+}$ ions at almost constant bulk sulfate concentration, as
summarized in Table~\ref{table2}.

A theoretical phase boundary can be represented by a functional relationship,
$\bar{\mu}_{\rm S}^{\rm peak}
= f \left( \bar{\mu}_{\rm C}^{\rm peak} \right)$.
Combining this with Eq.~(\ref{chempo}) and differentiating with
respect to ln[C], we obtain
\begin{equation}
\label{eqZZ}
z_{\rm C} - \left( f_i' \right)^{-1} z_{\rm S} = \frac{RT}{F}
\left( \frac{{\rm d} E_i^{\rm peak}}{{\rm d} \ln [ {\rm C} ]} \right)^{-1} \;,
\end{equation}
where $f'$ is the derivative of $f$ with respect to its argument, and the
subscripts identify the particular peak in question.
In the parameter region of interest, both the theoretical phase boundary and
the experimental dependence of the peak potentials on ln[C] are reasonably
approximated by straight lines. Therefore, combining the results for both
CV peaks, one obtains two linear equations for the two unknown quantities,
$z_{\rm C}$ and $z_{\rm S}$. Using our simulated data,
$\left( f_1' \right)^{-1} = -$0.52$\pm$0.03 and
$\left( f_2' \right)^{-1} = +$0.66$\pm$0.07,
together with the experimentally obtained
$\left( {{\rm d} E_1^{\rm peak}}/{{\rm d} \ln [ {\rm C} ]} \right)^{-1}
= +$42$\pm$3~V$^{-1}$
and
$\left( {{\rm d} E_2^{\rm peak}}/{{\rm d} \ln [ {\rm C} ]} \right)^{-1}
= +$94$\pm$6~V$^{-1}$,
we obtain
$z_{\rm C} = +$1.68$\pm$0.09 and $z_{\rm S} = -$1.14$\pm$0.16, where the
uncertainties only include statistical errors.
(If, instead, we had used the theoretical ground-state values,
$\left( f_1' \right)^{-1} = -$0.5 and
$\left( f_2' \right)^{-1} = +$1.0, obtained by equating the ground-state
energies in Table.~\ref{t2}, we would have obtained
$z_{\rm C} = +$1.54$\pm$0.07 and $z_{\rm S} = -$0.90$\pm$0.12.)

These values of $z_{\rm C}$ and $z_{\rm S}$ should be considered as
averages of weakly potential- and concentration-dependent electrovalences
in the region of experimental interest. To within the error bars they
agree with the values recently obtained by Shi {\it et al.\/}
\cite{ZSHI94C,ZSHI95B} by electrocapillary techniques.

\subsection{Further Parameter Refinement}
\label{secref}

In principle, Steps {\bf B} and {\bf C}
above can be iterated to further improve
the agreement between theory and experiment and correspondingly obtain
improved parameter estimates. However, as shown in the following sections
our simulated and experimentally measured data agree to a quite satisfactory
degree, using the lattice-gas parameters obtained after one iteration.
Since the iteration process is expensive,
both in terms of human and computer time, we therefore
terminated the loop after the first pass through Step {\bf C}.
Our final MC simulations were performed with the effective interactions
listed in Fig.~\ref{pic1}, using $z_{\rm C}$=+1.70 and $z_{\rm S}$=$-$1.15.

\section{Comparison of Simulated and Experimental Results}
\label{secres}

The room-temperature potential-scan path corresponding to the electrolyte
composition used in our experiments, was chosen such that the theoretical CV
peak separation along the scan equals the peak separation in
the experimental data shown in Figs.~\ref{eclr}(b) and~\ref{cvse}(a).
This scan is indicated by the dotted line labeled ``1" in Fig.~\ref{gs}.
The simulated CV current density [Eq.~(\ref{fd})]
along this potential scan is shown together
with our experimentally measured current
(positive-going scan) in Fig.~\ref{cvse}(a), and the corresponding integrated
charge densities [Eq.~(\ref{eqQ})] are shown in Fig.~\ref{cvse}(b).
As we pointed out in Sec.~\ref{sex}, there is previous experimental evidence
that at the negative end of the UPD potential range, sulfate
adsorbs as a submonolayer on top of the monolayer of Cu, with a
coverage $\Theta_{\rm S}^{(2)}\!\approx\! 0.2$ \cite{ZSHI94C}. These results
agree fully with our AES results, which are shown in Fig.~\ref{cg}.
The corresponding value of
$\alpha=0.6$ in Eq.~(\ref{eqQ}) was used to obtain the CV
current and surface-charge densities shown in Fig.~\ref{cvse}
and the sulfate coverage shown in Fig.~\ref{cg}
from the lattice-gas simulations.

With the aid of the ground-state diagram in Fig.~\ref{gs}
it is easy to analyse the simulated and experimental
results shown in Figs.~\ref{cvse} and \ref{cg}.
Starting from the negative end, we scan in the
direction of positive electrode potential ({\it i.e.\/}, from upper left
to lower right in Fig.~\ref{gs}). Near the CV peak at
approximately 200~mV (Peak~\#2),
sulfate begins to compete with Cu for the gold surface sites, resulting
in a third of the Cu desorbing and being replaced by
sulfate over a potential range of about 30~mV. Due to the strong effective
attraction between the sulfate and Cu adparticles, $\Phi_{\rm SC}^{(1)}$,
a mixed $(\sqrt{3}\!\times\!\sqrt{3})_{2/3}^{1/3}$ phase is formed, which
extends through the entire potential region between the two CV peaks.
The $(\sqrt{3}\!\times\!\sqrt{7})_{4/5}^{1/5}$ phase
which occurs in a narrow band in the ground-state
diagram between this phase and the (1$\times$1)$_1^0$ phase that
corresponds to the full Cu monolayer, is
not sufficiently stable to be observed at room temperature.

As the CV peak at approximately 300~mV (Peak~\#1) is reached,
most of the copper is desorbed within
a potential range of 10~mV. As it is thus deprived of the stabilizing influence
of the coadsorbed Cu, the sulfate is partly desorbed, reducing
$\Theta_{\rm S}$ from 1/3 to approximately 0.05.
The $(\sqrt{3}\!\times\!\sqrt{7})_0^{1/5}$ phase in the potential region
near 420~mV is consistent with experimental observations on copper-free
systems \cite{MAGN90,EDEN94}.
Eventually, more positive electrode potentials
cause the sulfate to form its saturated
$(\sqrt{3}\!\times\!\sqrt{3})_0^{1/3}$ hard-hexagon phase near 500~mV.
Since this phase has not been experimentally observed in {\it in situ\/}
experiments for this system, this is probably an artifact
of the model at very positive electrode potentials.

The scenario described here corresponds closely to that
proposed by HB \cite{HUCK91,HUCK92,BLUM94A,BLUM94B},
and the structure proposed for
the mixed $(\sqrt{3}\!\times\!\sqrt{3})_{2/3}^{1/3}$ phase
between the two CV peaks has recently been strongly supported by
Toney {\it et al.\/} with {\it in situ\/}
surface X-ray scattering \cite{TONE95}.
The nearly linear dependence of the two coverages on each other
for $\Theta_{\rm C}$ between approximately 0.1 and 0.6, together with
the broad maximum of the differential
coadsorption ratio,
[d$(\Theta_{\rm S} \! + \! \Theta_{\rm S}^{(2)})$/d$\Theta_{\rm C}$
is close to 0.5 in this region] are shown in Fig.~\ref{tatb3}.
(Also compare Fig.~\ref{tatb3}(a) with the insert in Fig.~4(b) of
Ref.~\cite{ZSHI94C}.)
These results show that this system provides an example of the
enhanced-adsorption phenomenon described by Rikvold and Deakin \cite{RIKV91A}.

Very close to the phase transitions, the CV current profile may be
significantly distorted from the equilibrium result represented by
Eq.~(\ref{fd}) due to kinetic effects. In addition, the transition
may be rounded by defects or impurities on the adsorbent surface, which can
have significant effects even at very low concentrations \cite{PERE93}.
As a result of these effects, the simulated
quantities that are most reliably compared with experimental results
are the CV currents some distance away from their maxima, and the corresponding
charge densities. With this caveat, we find the agreement between our
simulations and experiments quite good. The shape of the foot region of
Peak~\#2 is well reproduced [Fig.~\ref{cvse}(a)], and the simulated
charge transfer across that peak agrees with the experimental value to
within about 15\% [Fig.~\ref{cvse}(b)]. Likewise, good agreement is found
with the Cu and sulfate coverages measured by AES [Fig.~\ref{cg}].
The background current in
the ordered-phase region between the peaks also shows excellent
agreement [Fig.~\ref{cvse}(a)].
Somewhat less satisfactory agreement is obtained for Peak~\#1. The simulated
peak is somewhat too narrow at the base, and the foot on the
positive-potential side of the peak, although visible, is lower
than its experimental counterpart.
We find these discrepancies between the simulated and observed current
densities (which were recorded at a potential scan rate of 2~mV/s and show
noticeable hysteresis) to be within what one should expect from our
essentially equilibrium
calculation, even if the effective lattice-gas interaction
were highly accurate. We therefore believe that
a study of the effects of kinetics and defects in
our model would be very valuable.

To investigate the effects of changing the electrolyte concentration in the
lattice-gas model,
two additional isothermal potential-scan
paths were simulated. We assumed the same bulk
sulfate concentration as for Path~1
and changed $\bar{\mu}_{\rm C}$
[Eq.~(\ref{chempo})] to simulate bulk copper concentrations
of 5~mM (Path~2) and 0.2~mM (Path~3), respectively.
All three paths are shown in Fig.~\ref{gs} as dotted straight lines.
Simulations of current densities and coverages
corresponding to Paths~2 and~3 are shown in Figs.~\ref{dipath}(a)
and~\ref{dipath}(b), respectively.
Along Path~2, which corresponds to the highest sulfate concentration,
a wider potential interval is required to
traverse the mixed-phase region; hence
the CV peak separation is larger than for Path~1. Path~3,
conversely, corresponds to the lowest sulfate
concentration. It exits quickly from the
$(\sqrt{3} \! \times \! \sqrt{3})_{2/3}^{1/3}$
phase into the disordered low-coverage phase,
giving a CV peak separation that is smaller than
for the other two paths, as well as lower coverages of both Cu and sulfate
immediately to the positive-potential side of Peak~\#1.
We also note that the experimentally observed asymmetric shape of the
foot of Peak~\#1 is enhanced for lower bulk copper concentrations,
as is particularly evident from the CV current along Path~3, shown in
Fig.~\ref{dipath}(a).

\section{Finite-Size Scaling Analysis}
\label{secFSS}

The model studied here is equivalent to a generalized
triangular-lattice Blume-Emery-Griffiths (BEG) \cite{BEG}
model with interactions of longer than nearest-neighbor range. The
lattice-gas Hamiltonian in Eq.~(\ref{eq1}) can be transformed into that
of the equivalent BEG model by the following transformation,
\begin{mathletters}
\begin{eqnarray}
M &=& \Theta_{\rm C} - \Theta_{\rm S}\\
Q &=& \Theta_{\rm C} + \Theta_{\rm S}\\
H &=& \frac{1}{2} \left( \bar{\mu}_{\rm C} - \bar{\mu}_{\rm S}  \right)\\
D &=& - \frac{1}{2} \left( \bar{\mu}_{\rm C} + \bar{\mu}_{\rm S}  \right)
\; ,
\label{eqBEG}
\end{eqnarray}
\end{mathletters}
where $M$ is the density conjugate to $H$, and $Q$ is the density
conjugate to $-$$D$.

By comparing the topology of the
ground-state diagram shown in Fig.~\ref{gs} with those
for the generalized BEG models illustrated in
Refs.~\cite{RIKV88B,COLL89,RIKV91A,RIKV91B,RIKV92,COLL88,KIME92},
one can deduce the order and universality class of the phase transitions
that correspond to CV Peak~\#1 and~\#2.

The transition at Peak~\#1 corresponds to the phase transition seen
in the BEG
models as $D$ is increased for small
nonzero values of $H$ and $T$. This is a
first-order transition in which the total coverage $Q$ drops
discontinuously from near one to near zero. As is generally the case for
first-order transitions \cite{BIND92}, the adsorbate heat capacity $C$ and
the nonordering response functions, ${\partial M}/{\partial H}$
and ${\partial Q}/{\partial D}$, diverge proportionally with the
system area, $L^2$. This divergence is an expression of the discontinuities
in $M$ and $Q$ (or equivalently in $\Theta_{\rm C}$ and $\Theta_{\rm S}$) and
the latent heat associated with a first-order transition.

The transition at Peak~\#2 corresponds to the phase transition seen
in the BEG models as $H$ is increased for low $T$ and negative $D$.
This is a second-order transition which for
a system on a triangular lattice is in the same universality class as
the hard-hexagon model, namely that of the three-state Potts model
\cite{DOMA78}. The standard critical exponents
in this universality class
are $\nu \! = \! \frac{5}{6}$ for the correlation length,
$\alpha \! = \! \frac{1}{3}$ for the heat capacity, and
$\beta \! = \! \frac{1}{9}$ for the order parameter \cite{BAXT82B}.
At a second-order phase transition the heat capacity diverges with the
system size as $L^{\alpha / \nu}$ \cite{BIND92},
which for this universality
class becomes $L^{2/5}$. The nonordering response functions are coupled
to the heat capacity and are expected to share the same
divergence \cite{RIKV85}.

We have checked these predictions by MC simulations for systems
with $L$=15, 30, and~45. The results are shown
for Peak~\#1 {\it vs.\/} $L^2$ in
Fig.~\ref{figFSS}(a)
and
for Peak~\#2 {\it vs.\/} $L^{2/5}$ in
Fig.~\ref{figFSS}(b).
(To reduce the effects of the finite-size rounding of the phase
transitions, we show only the scaling results for the response functions
corresponding to a direction pointing away from the transition line
in the phase diagram,
$-$$RT(\partial Q / \partial D)$ for Peak~\#1
and
$RT(\partial M / \partial H)$ for Peak~\#2.
In both cases the response function not shown has the same scaling
behavior as the one shown, but its magnitude is much smaller.)
The results shown in Fig.~\ref{figFSS} are consistent with the scaling
relations predicted above.

\section{Discussion and Conclusions}
\label{secD}

In this study we have combined a computational and theoretical
lattice-gas modeling approach with electrochemical and UHV experiments
to study the UPD of Cu on Au(111) single-crystal electrodes in the
presence of sulfate. In agreement with other experimental and theoretical
studies we find that in the potential region between the two sharp CV peaks,
the electrode is covered by a mixed adlayer of
$(\sqrt{3}$$\times$$\sqrt{3})R30^{\circ}$ symmetry, consisting of 2/3~ML
Cu and 1/3~ML sulfate. This ordered-phase region is limited on the
positive-potential side by a
first-order phase transition to a disordered low-coverage
phase, followed at still higher potentials by transitions to pure sulfate
phases. On the negative-potential side the mixed phase terminates at a
second-order phase
transition to a full monolayer of Cu. The mixed phase is stabilized by
relatively strong, effective attractive interactions between
nearest-neighbor copper and
sulfate adparticles.

The lattice-gas parameters were obtained by a three-step procedure.
First, a group-theoretical ground-state calculation was used to determine the
intervals
of interaction energies which produce ordered phases that agree with
the experimental observations.
Second, MC simulations were used to produce adsorbate coverages
and CV current profiles at room temperature, and the interaction energies were
adjusted until optimal agreement with all the available experimental data was
obtained.
Third, the effective electrovalences were adjusted to obtain agreement between
the simulations and experiments regarding
the dependences of the CV peak potentials on the bulk copper concentration.
Our final estimates for the effective lateral interactions are given in
Fig.~\ref{pic1}. It is difficult to attach error estimates to these
values, and there is no guarantee that even better overall agreement
could not be obtained with a different parameter set. Seen as a formal
optimization problem, the parameter adjustment we have performed here by the
procedure detailed in Subsecs.~\ref{secgs}--D, is a daunting task.
We are currently studying strategies
that could, at least to some extent, improve
the reliability of the parameter optimization procedure with a manageable
computational effort \cite{ASSE95}. However,
no first-principles method exists that could calculate the effective
lattice-gas parameters in this complicated, metal--water--multi-ion system
with the degree of accuracy needed to calculate thermodynamic
and structural quantities for
quantitative comparison with experiments \cite{FEIB89,EINS95}. We are
of the opinion that our approach represents the only method for
determining effective interaction energies in complicated electrochemical
systems, which is practically feasible at the present stage of theoretical
and computational development.

Despite the cautionary note sounded in the preceding paragraph, we point out
that our estimated interactions appear entirely sensible from a chemical
point of view. Two aspects of the interactions are particularly interesting.
First, the effective nearest-neighbor sulfate-Cu attraction is
relatively strong:
$\Phi_{\rm SC}^{(1)} \! = \!$+4.0~kJ/mol. This is consistent with the short
Cu-O bond length (2.15~{\AA})
found in recent {\it in situ\/} X-ray diffraction measurements \cite{TONE95}.
Second, and perhaps more surprisingly: the nearest-neighbor Cu-Cu
interaction is moderately repulsive:
$\Phi_{\rm CC}^{(1)} \! = \!$$-$1.8~kJ/mol. However,
interaction energies on surfaces are usually much weaker than the
interactions between the same atoms in the bulk solid,
and the presence of the fluid and residual charges on the adsorbed particles
may further modify the effective interactions to turn attraction into
repulsion \cite{KOLB78}.
For the present system, the conclusion that nearest-neighbor Cu
adparticles interact repulsively with each other,
was already reached by Magnussen {\it et al.\/} \cite{MAGN91}. We consider it
a measure of the robustness of our parameter-estimation
procedure that we arrived at the repulsive
nature of this effective interaction independently and only after lengthy
attempts at obtaining a reasonable shape for Peak~\#2 with an
attractive nearest-neighbor interaction had failed.

The residual disagreement between our simulated and experimental
results near the CV current
peaks most likely arise from defects on the electrode surface, and
from the use of equilibrium
statistical mechanics to study in detail a phenomenon that has significant
kinetic aspects. We believe that the effects of defects and kinetics in our
model present promising research topics, and we hope to return to their study
in the future.

\acknowledgements

%We gratefully acknowledge useful discussions with L.~Blum.
Helpful comments on the manuscript were provided by
M.~A.\ Novotny, R.~A.\ Ramos, and S.~W.\ Sides.
P.~A.~R.\ appreciates the hospitality and support of the Centre for the
Physics of Materials and the Department of Physics of McGill University
during the final stages of this work.
This research was supported by Florida State University
through the Supercomputer Computations Research Institute
(DOE Contract No.\ DE-FC05-85-ER25000) and the Center
for Materials Research and Technology, and by NSF grant No.~DMR-9315969.
Work at The University of Illinois was supported by
NSF grant No.~CHE-9411184 and by the Frederick Seitz
Materials Research Laboratory under DOE Contract No.\ DE-AC02-76-ER01198.

\newpage

%\bibliography{/a/alpha2/home/scri42/users/rikvold%
%/decstation/tex/biblio/elchem}
%\bibliographystyle{prsty}

%\bibliography{/a/alpha2/home/scri42/users/rikvold%
%/decstation/tex/biblio/elchem}
%\bibliographystyle{unsrt}

\newpage

%%%%%%%%%%%%%%%%Tables%%%%%%%%%%%%%%%%

\begin{table}
\caption[]{
Representative experimental conditions employed in the
present paper and in some other voltammetric studies.
The peak separations are calculated as averages over positive- and
negative-going potential scans
and are taken from figures or tables in the cited references.
}
\begin{tabular}{llll}
Reference & Electrolyte & Peak separation & Scan rate \\ \hline
This work & 1~mM CuSO$_4$ + 0.1~mM H$_2$SO$_4$ & 100 mV  & 2 mV/s \\
Kolb {\it et al.\/} \cite{KOLB86,KOLB87}
                     &1~mM CuSO$_4$ + 50~mM H$_2$SO$_4$ & 160 mV    & 1 mV/s \\
Schultze {\it et al.\/} \cite{SCHU76} & 1 M HClO$_4$ + 1 mM Cu$^{2+}$
+ 1 mM SO$_4^{2-}$ &
130 mV & 20 mV/s \\ Shi {\it et al.\/}
\cite{ZSHI94B,ZSHI94C} &1 mM Cu(ClO$_4)_2$ + 5 mM K$_2$SO$_4$ &
140 mV&5 mV/s \\
           &+ 0.1 M HClO$_4$    &     &    \\
Omar {\it et al.\/} \cite{OMAR93} & 5 mM CuSO$_4$ + 90 mM H$_2$SO$_4$ &
 150 mV  & 5 mV/s \\
\end{tabular}
\label{table1}
\end{table}

\newpage

\begin{table}
\caption[]{List of ordered phases
($X\!\times\!Y$)$_{\Theta_{\rm C}}^{\Theta_{\rm S}}$
and their ground-state energies per site.
Only those phases are included
that are realized as ground states for interactions in the region of
experimental interest.}
\[
\begin{array}{lll}
& \mbox{\normalsize ${\rm Phase}$} &
\mbox{\normalsize ${\rm Ground\!\!-\!\!state \;energy \; per \;
site}$} \\*[0.06in]
\hline
1 & (1\!\times\! 1)^0_0 & 0 \\
2 & (1\!\times\! 1)^0_1 & -\bar{\mu}_{\rm C}-3\Phi_{\rm
CC}^{(1)}-3\Phi_{\rm CC}^{(2)} \\
3 & (\sqrt{3}\!\times\! \sqrt{3})^{1/3}_{2/3} &
-(\bar{\mu}_{\rm S}+2\bar{\mu}_{\rm C})/3
-\Phi_{\rm CC}^{(1)}-2\Phi_{\rm CC}^{(2)}-2\Phi_{\rm SC}^{(1)}
-\Phi_{\rm SS}^{(2)}-2\Phi_{\rm SS}^{(t)}/3 \\
4 & (\sqrt{3}\!\times\! \sqrt{3})^{1/3}_{1/3} &
-(\bar{\mu}_{\rm S}+\bar{\mu}_{\rm C})/3
-\Phi_{\rm CC}^{(2)}
-\Phi_{\rm SC}^{(1)}
-\Phi_{\rm SS}^{(2)}-2\Phi_{\rm SS}^{(t)}/3  \\
5 & (\sqrt{3}\!\times\! \sqrt{3})^{0}_{2/3} & -2\bar{\mu}_{\rm C}/3
-\Phi_{\rm CC}^{(1)}-2\Phi_{\rm CC}^{(2)}  \\
6 & (\sqrt{3}\!\times\! \sqrt{3})^{1/3}_{0} & -\bar{\mu}_{\rm S}/3
-\Phi_{\rm SS}^{(2)}-2\Phi_{\rm SS}^{(t)}/3 \\
7 & (\sqrt{3}\!\times\! \sqrt{3})^{0}_{1/3} & -\bar{\mu}_{\rm
C}/3 -\Phi_{\rm CC}^{(2)} \\
%8X & (\sqrt{7}\!\times\! \sqrt{7})^{1/7}_{6/7} &
%-(\bar{\mu}_{\rm S}+6\bar{\mu}_{\rm C}+15\Phi_{\rm CC}^{(1)}
%+15\Phi_{\rm CC}^{(2)}
%+6\Phi_{\rm SC}^{(1)}+3\Phi_{\rm SS}^{(4)})/7 \\
8 & (\sqrt{3}\!\times\!\sqrt{7})_{4/5}^{1/5} & -(\bar{\mu}_{\rm S}
+4\bar{\mu}_{\rm C} + \Phi_{\rm SS}^{(2)}+2\Phi_{\rm SS}^{(4)}
+9\Phi_{\rm CC}^{(1)}+10\Phi_{\rm CC}^{(2)}+6\Phi_{\rm SC}^{(1)})/5 \\
9 & (\sqrt{7}\!\times\! \sqrt{7})^{1/7}_{0} &
-(\bar{\mu}_{\rm S}+3\Phi_{\rm SS}^{(4)})/7 \\
%10X & (3\!\times\!3)^{2/9}_{7/9} &
%-(2\bar{\mu}_{\rm S}+7\bar{\mu}_{\rm C}+15\Phi_{\rm CC}^{(1)}
%+18\Phi_{\rm CC}^{(2)}+12\Phi_{\rm SC}^{(1)}+3\Phi_{\rm SS}^{(2)})/9 \\
10 & (\sqrt{3}\!\times\! \sqrt{7})^{1/5}_{0} & -(\bar{\mu}_{\rm S}
+\Phi_{\rm SS}^{(2)}+2\Phi_{\rm SS}^{(4)})/5 \\
\hline
\end{array}
\]
\label{t2}
\end{table}

\newpage

\begin{table}
\caption[]{
CV peak positions and separations for different electrolyte compositions
at a potential scan rate of 10~mV/s, from Ref.~\cite{OMAR93}.
The compositions used were $x$~mM CuSO$_4$ + 90~mM H$_2$SO$_4$.
The peak positions given in the table are in mV {\it vs.\/}
the Normal Hydrogen
Electrode. They are averages over the positive- and negative-going scans
in Fig.~4 of Ref.~\cite{OMAR93}, taken from Table~2 of Ref.~\cite{BLUM94A}.
These results are used in calculating the effective electrovalences in
Subsec.~\protect\ref{secZZ}.
}
\begin{tabular}{llll}
$x$ & Peak~\#1 & Peak~\#2  & Peak separation \\ \hline
0.5 & 444~mV & 303~mV & 141~mV \\
5.0 & 496~mV & 327~mV & 169~mV \\
50  & 554~mV & 352~mV & 202~mV \\
\end{tabular}
\label{table2}
\end{table}

\newpage
%%%%%%%%%%%%%%%%Figure Captions%%%%%%%%%%%%%%%%

\begin{figure}
\caption[]{
(a) Our experimental CV current densities for a Au(111) electrode in
0.1~mM H$_2$SO$_4$ at a potential scan rate of 10~mV/s (dotted curve)
and in 1~mM CuSO$_4$ + 0.1~mM H$_2$SO$_4$ at 2~mV/s (solid and
dash-dotted curves). The differences between the latter two curves provide
a measure of the experimental uncertainties.
The arrows indicate the potential-scan directions.
Note that the current density is proportional to the scan rate.
Thus, the copper-free voltammogram corresponds to a very small charge
transfer per unit potential change, compared to the copper-containing case.
(b) Integrated charge density for the positive-going scan
with Cu$^{2+}$-containing electrolyte (solid curve).
[The corresponding current density is the dash-dotted curve in panel (a),
which for clarity is also shown against the
right-hand vertical axis in this panel.]
The charge density is defined as zero at +366~mV {\it vs.\/} Ag/AgCl.
}
\label{eclr}
\end{figure}

\begin{figure}
\caption[]{
Auger electron spectra at 3~keV of the Au(111) electrode covered
by uderpotentially deposited Cu and coadsorbed sulfate.  The
emersions were carried out at several electrode potentials,
as indicated {\it vs.\/} Ag/AgCl.
These data were used to calculate the Cu and sulfate coverages shown in
Figs.~\protect\ref{cg} and~\protect\ref{tatb3}.
}
\label{figA2}
\end{figure}

\begin{figure}
\caption[]{
The ($\sqrt 3$$\times$$\sqrt 3$)R30$^\circ$
LEED pattern at 91.1 eV electron energy, obtained from the
Au(111) electrode emersed from the 1~mM CuSO$_4$ +
0.1~mM H$_2$SO$_4$ solution at $E$=220~mV {\it vs.\/} Ag/AgCl,
which is in the potential region between the two CV peaks shown in
Fig.~\protect\ref{eclr}.
}
\label{figA1}
\end{figure}

\begin{figure}
\caption[]{
The relative positions of Cu ({\large $\bullet$}) and sulfate
($\triangle$) corresponding to the effective interactions in
Eq.~(\protect\ref{eq1}). The number underneath each bond representation is the
corresponding effective interaction energy
obtained in this work, given in kJ/mol.
In particular, the configuration
corresponding to $\Phi_{\rm SS}^{(1)}$,
in which the corners of the
sulfate triangles touch, is forbidden; hence the infinitely
strong ``hard-hexagon" repulsion.
The interactions are invariant under symmetry operations on the lattice.
}
\label{pic1}
\end{figure}

\begin{figure}
\caption[]{
The ground-state diagram obtained with the effective interaction constants
given in Fig.~\protect\ref{pic1}. The zero-temperature phase boundaries are
represented by the solid lines, and the phases are indicated as
($X\!\times\!Y$)$_{\Theta_{\rm C}}^{\Theta_{\rm S}}$. Superimposed on the
ground-state diagram are the positions of CV Peak~\#1 (filled diamonds)
and Peak~\#2 (filled squares) at room temperature.
The three
points marked $\bullet$ are low maxima associated with Cu adsorption
without coadsorbed sulfate. The potential scan path for
the simulation corresponding to our experiments with
the 1~mM CuSO$_4$ + 0.1~mM H$_2$SO$_4$ electrolyte is shown as the dotted
line labeled ``1,'' whereas the lines labeled ``2'' and ``3'' correspond to
additional simulations for 5~mM and 0.2~mM Cu$^{2+}$ with unchanged sulfate
concentration, respectively. For all three lines, the upper, left end point
corresponds to 120~mV {\it vs.\/} Ag/AgCl, and the lower, right end point
corresponds to 420~mV.
}
\label{gs}
\end{figure}

\begin{figure}
\caption[]{
(a)
Experimental (dot-dashed) and simulated (solid)
CV current densities. The experimental curve is the same as the positive-going
scan shown in Fig.~\protect\ref{eclr}(b). The left-hand vertical scale shows
the current density for a scan rate of 2~mV/s, whereas the right-hand scale
shows the current density
normalized by the scan rate, in units of elementary charges per
mV and Au(111) unit cell.
(b)
Experimental (dot-dashed) and simulated (solid) integrated charge
transfers. The experimental curve is the same as
shown in Fig.~\protect\ref{eclr}(b). The dashed and long-dashed curves
represent the simulated,
partial charge density corresponding to copper and sulfate, respectively.
The experimental charge density is defined as zero at +366~mV.
The data points marked $\times$ in both panel (a) and (b) represent
additional simulation results for $L$=45. They show that finite-size
effects in the simulations are negligible, except very close to the
maxima of the CV peaks.
}
\label{cvse}
\end{figure}

\begin{figure}
\caption[]{
Simulated coverages of Cu, $\Theta_C$ (dashed), first-layer sulfate,
$\Theta_S$ (dotted), and total sulfate,
$\Theta_S$+$\Theta_S^{(2)}$ (long-dashed), shown together with the
corresponding results from our AES experiments, $\Box$ and $\Diamond$,
respectively.
}
\label{cg}
\end{figure}

\begin{figure}
\caption[]{
(a) The total sulfate
coverage, $\Theta_{\rm S}$+$\Theta_{\rm S}^{(2)}$, shown {\it vs.\/}
$\Theta_{\rm C}$.
The data points shown as crossed error bars represent the AES data from
Fig.~\protect\ref{cg}.
(b) Finite-difference estimate of
the differential coadsorption ratio
d($\Theta_{\rm S} \! + \! \Theta_{\rm S}^{(2)}$)/d$\Theta_{\rm C}$
as a function of $\Theta_{\rm C}$.
The ratio exhibits a broad maximum,
corresponding to the linear section in panel (a), indicative of
mutually enhanced adsorption.
}
\label{tatb3}
\end{figure}

\begin{figure}
\caption[]{
Simulation results for Path~2 (5~mM Cu$^{2+}$) and Path~3 (0.2~mM Cu$^{2+}$)
in Fig.~\protect\ref{gs}.
(a)
CV current densities for Path~2 (dashed) and Path~3 (solid).
(b)
Coverages for Path~2
($\Theta_{\rm C}$: dashed;
$\Theta_{\rm S} \! + \! \Theta_{\rm S}^{(2)}$: dot-dashed)
and Path~3
($\Theta_{\rm C}$: solid;
$\Theta_{\rm S} \! + \! \Theta_{\rm S}^{(2)}$: dotted).
}
\label{dipath}
\end{figure}

\begin{figure}
\caption[]{
Finite-size scaling plots of the maximum values of the dimensionless
adsorbate heat capacity, $C/R$, and nonordering response functions,
$-$$RT(\partial Q / \partial D)$ and
$RT(\partial M / \partial H)$, for system sizes $L$=15, 30, and~45.
The straight lines are unweighted least-squares fits to the data points.
See details in Sec.~\protect\ref{secFSS}.
(a)
At Peak~\#1, showing $C/R$
($\Diamond$ and solid line)
and $-$$RT(\partial Q / \partial D)$
($\Box$ and dashed line)
{\it vs.\/} $L^2$. The response function has been divided by~10 for
easy visual comparison with the heat capacity.
(b)
At Peak~\#2, showing $C/R$
($\Diamond$ and solid line)
and $RT(\partial M / \partial H)$
($\bigcirc$ and dot-dashed line)
{\it vs.\/} $L^{2/5}$.
}
\label{figFSS}
\end{figure}

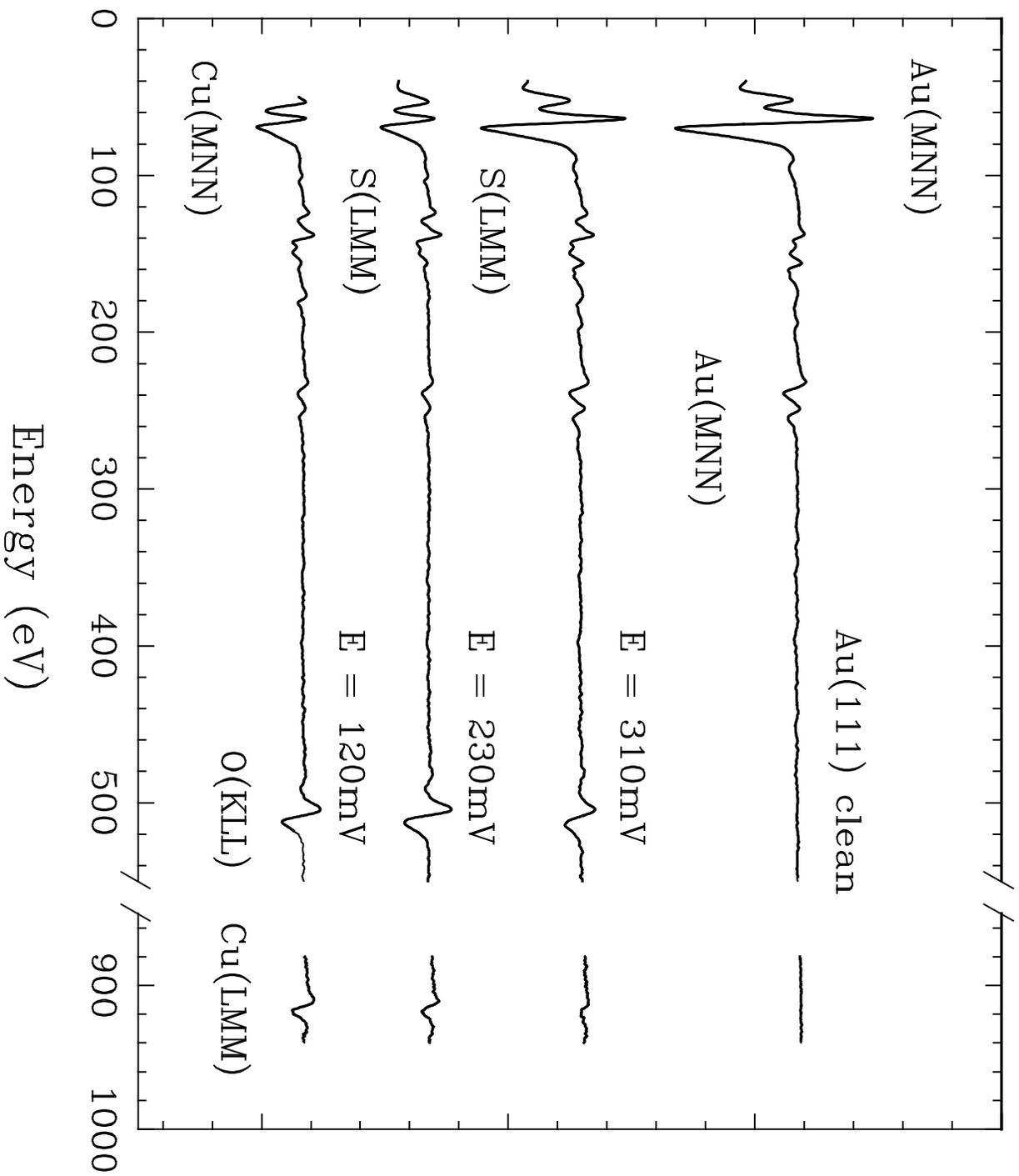
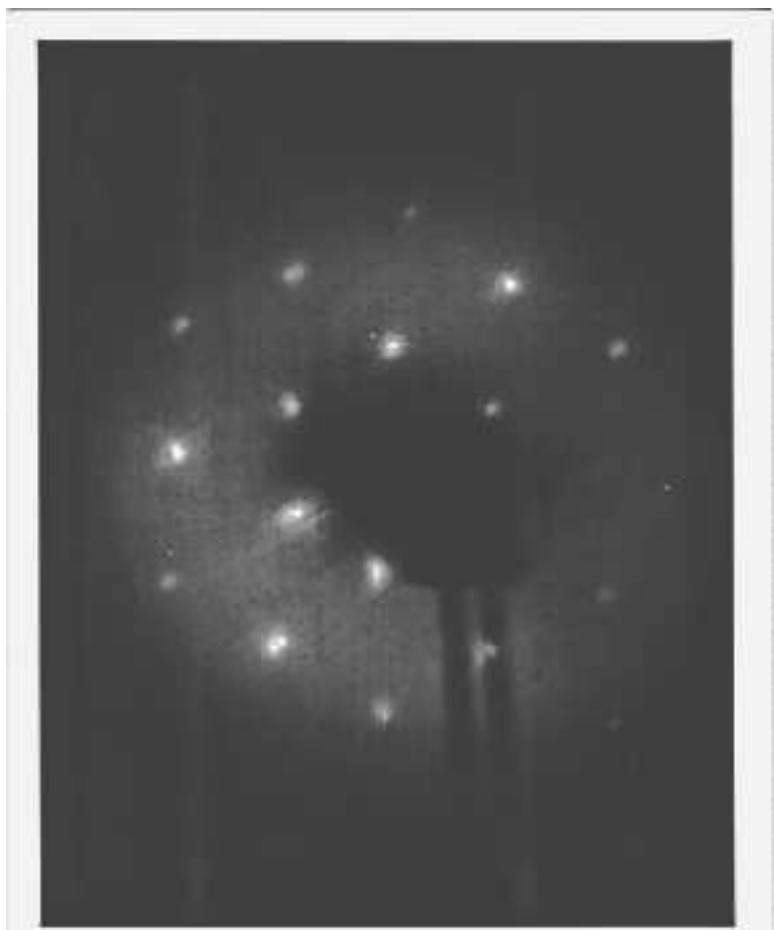
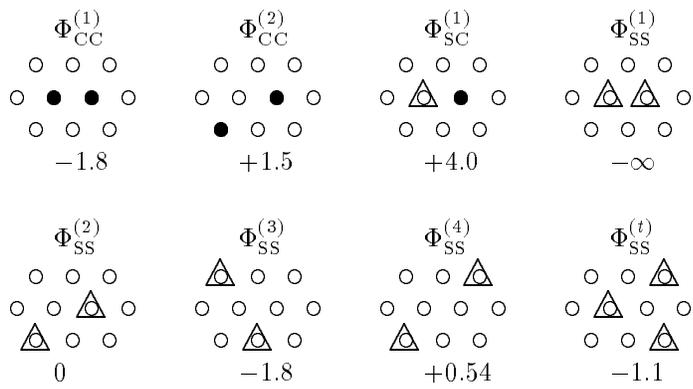
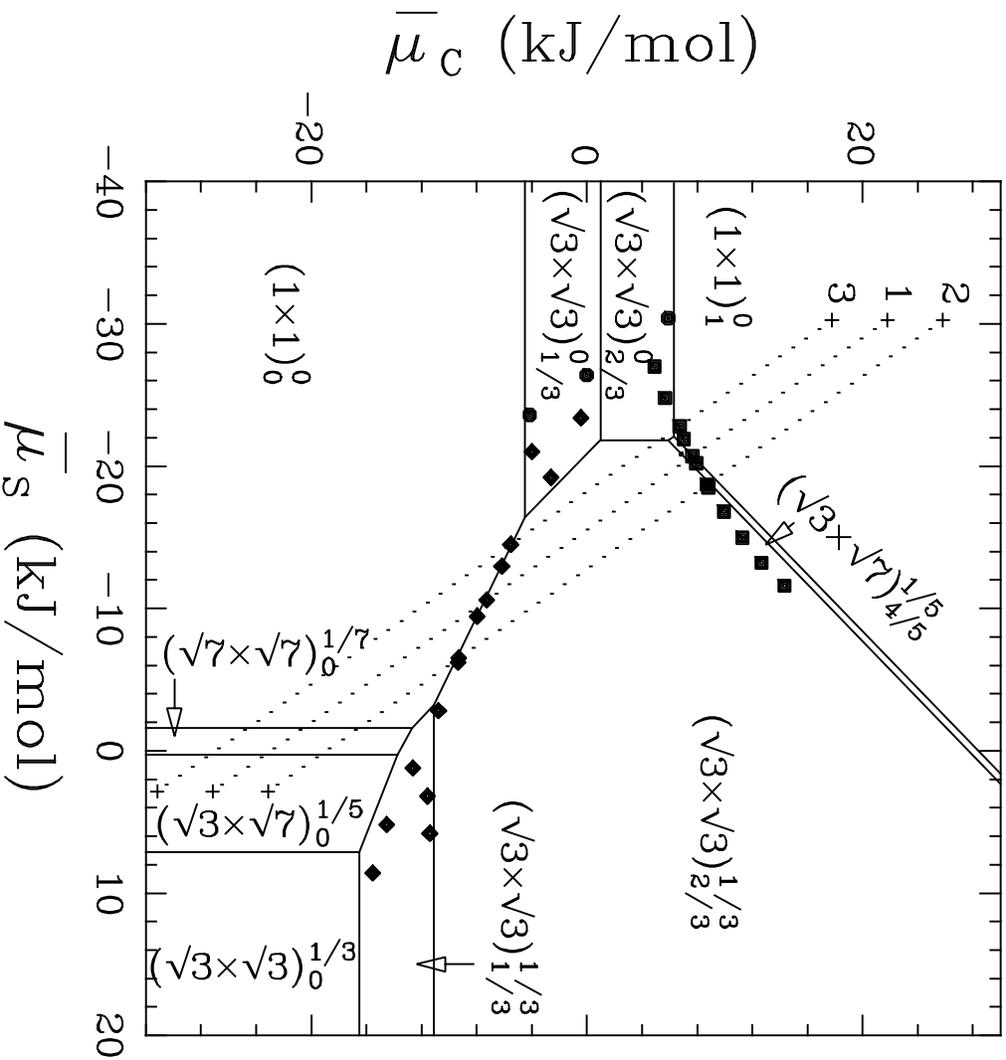
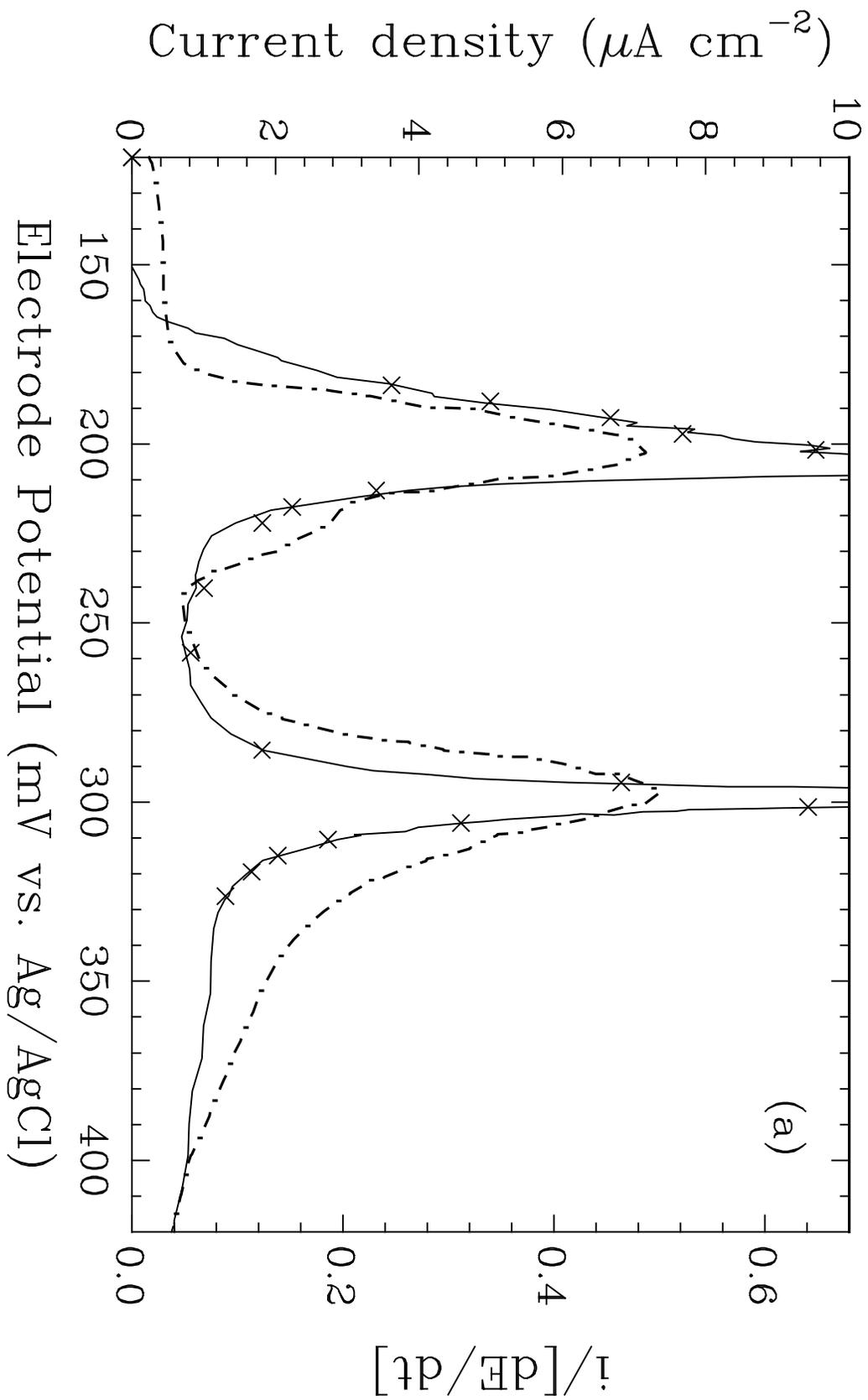
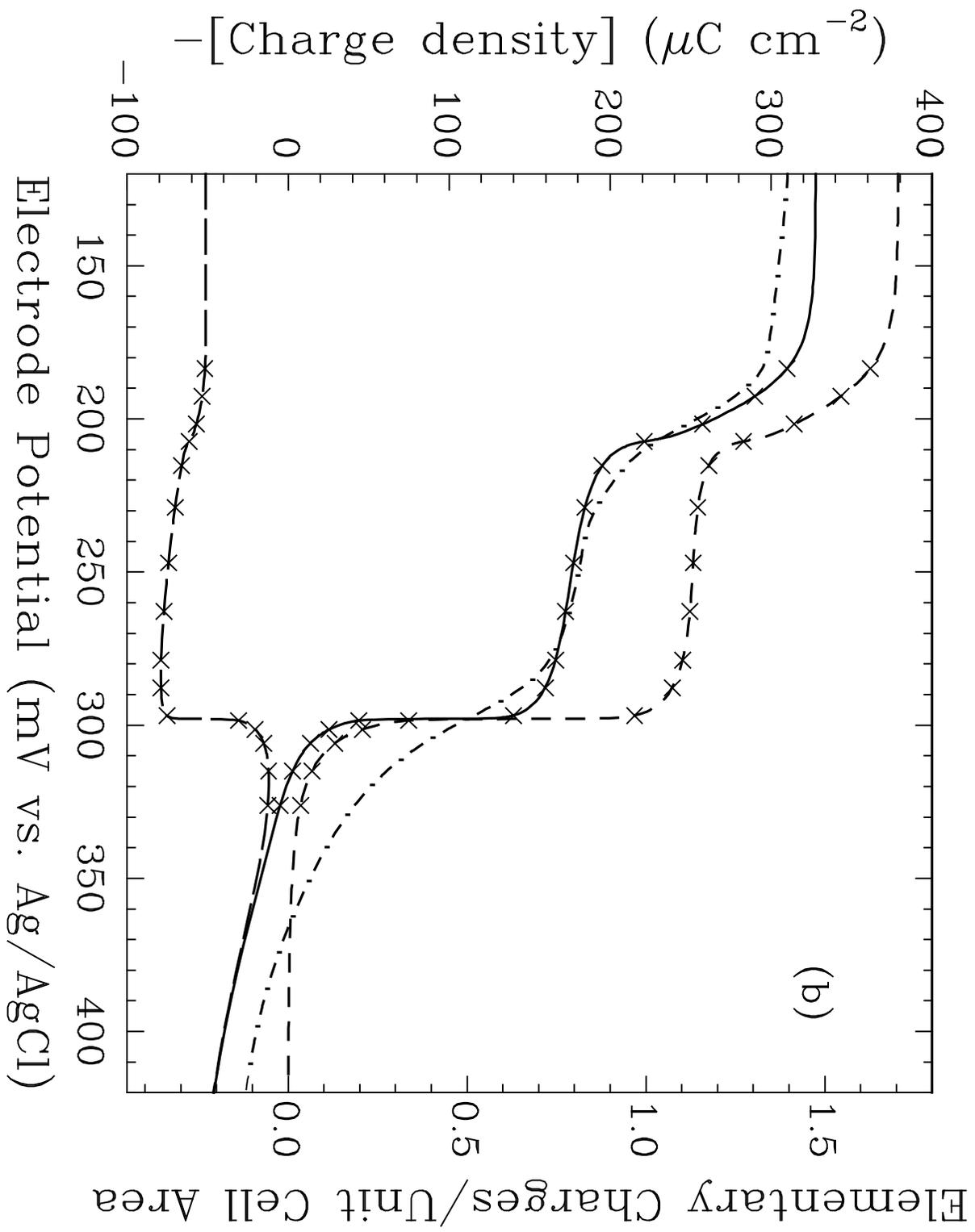
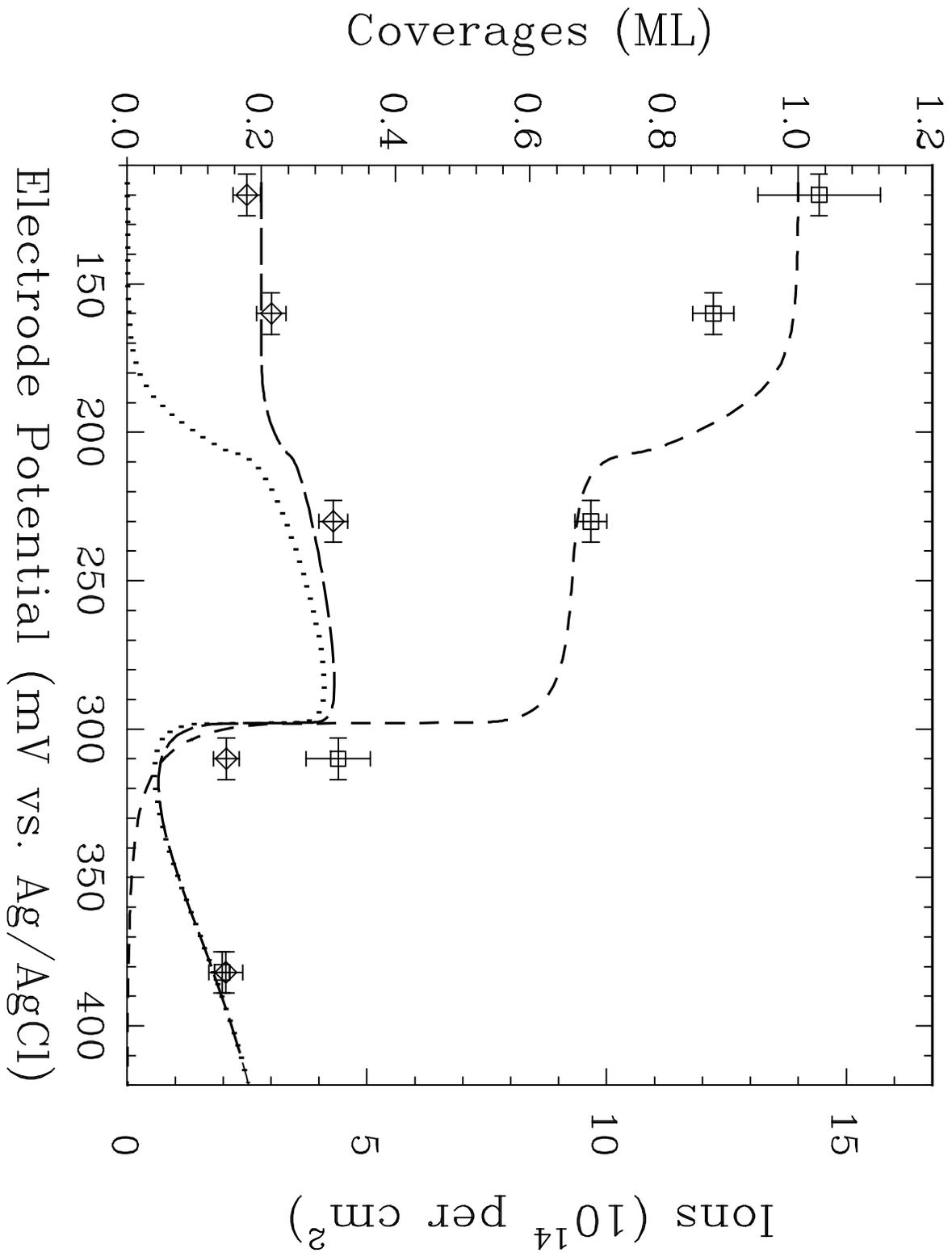
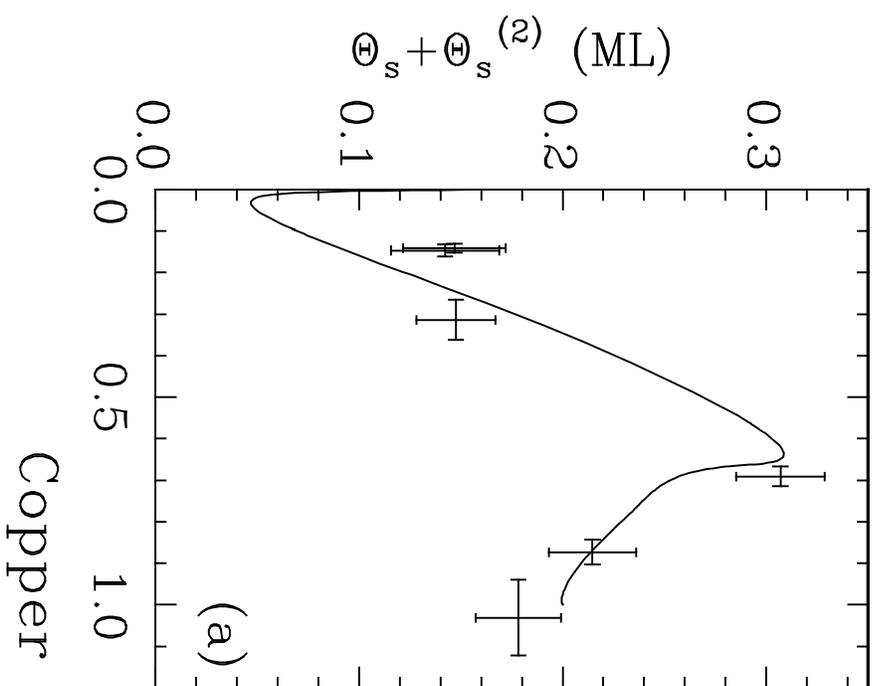
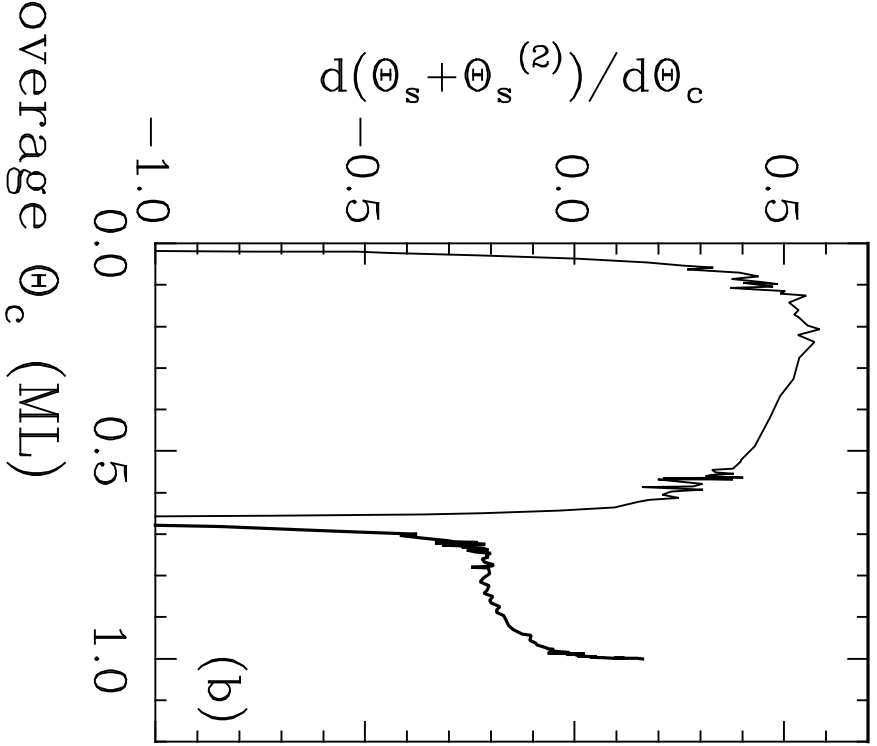
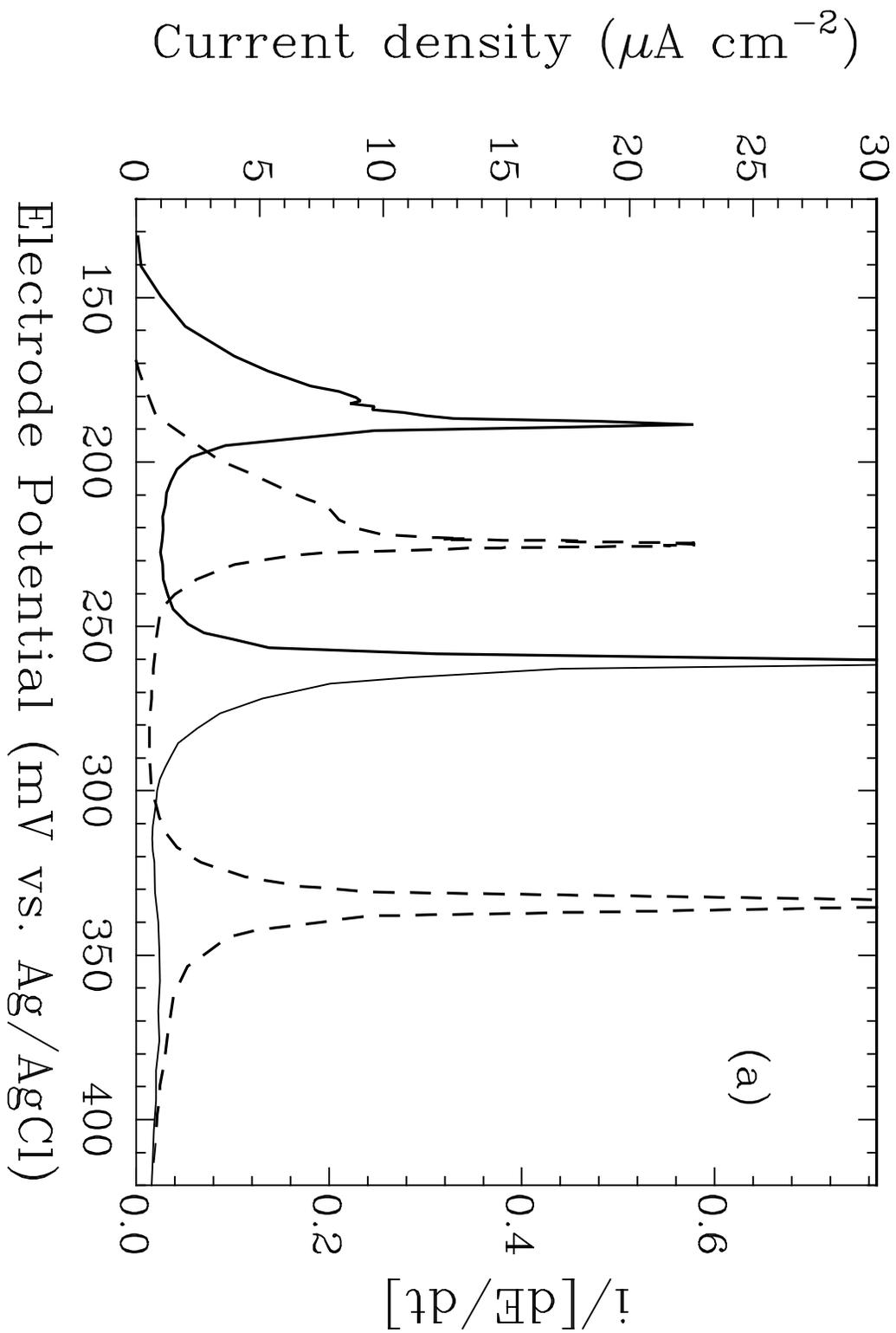
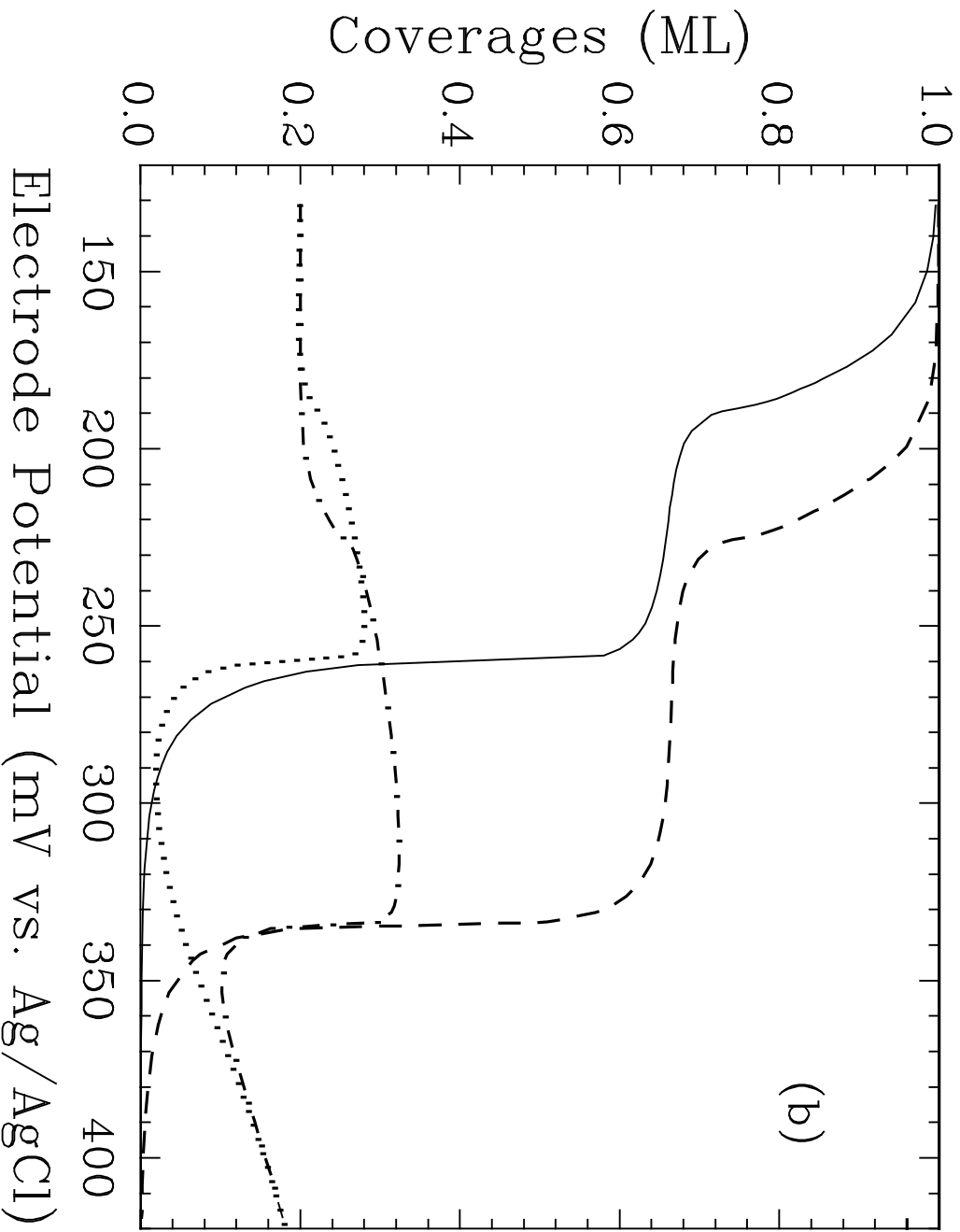
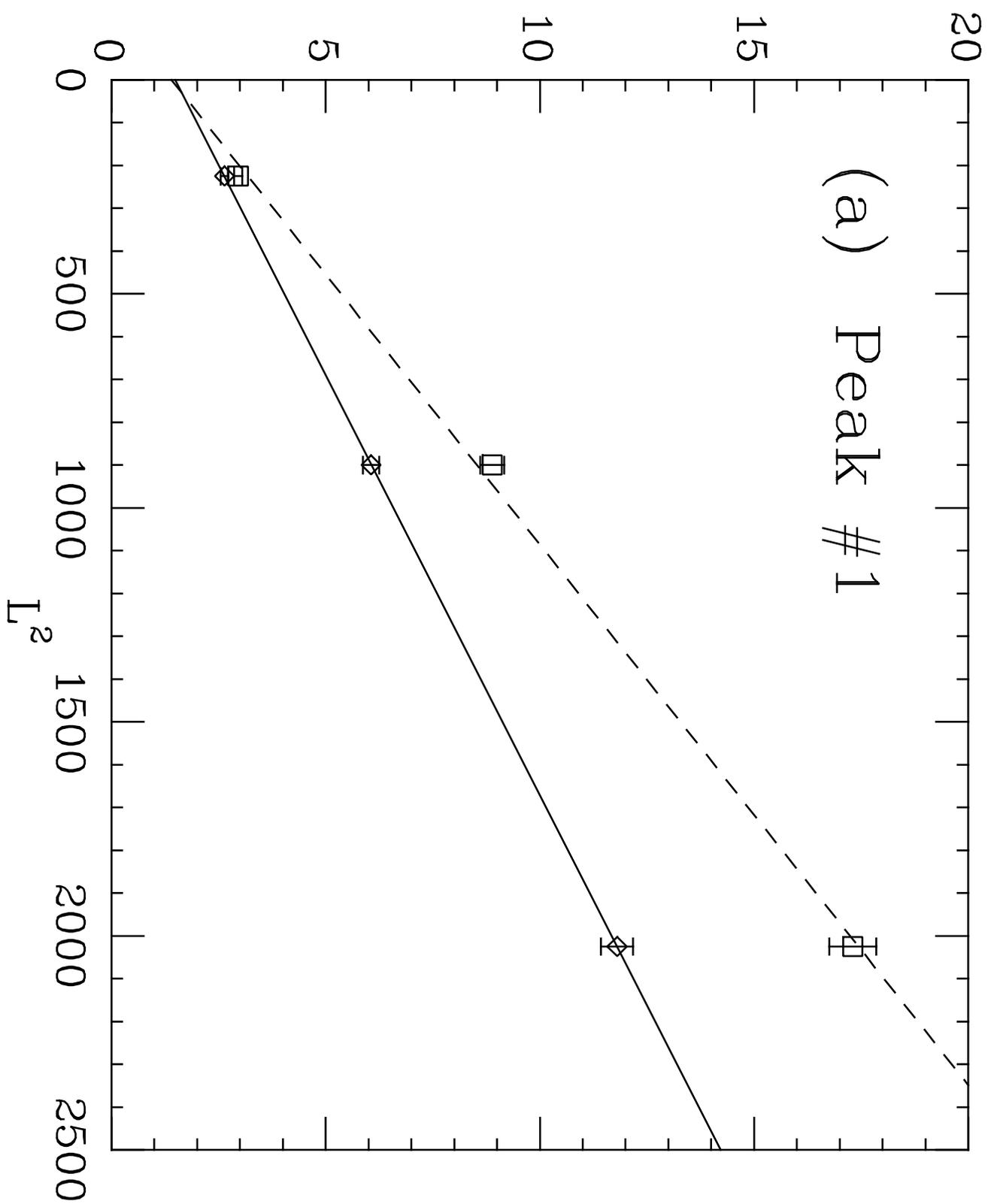
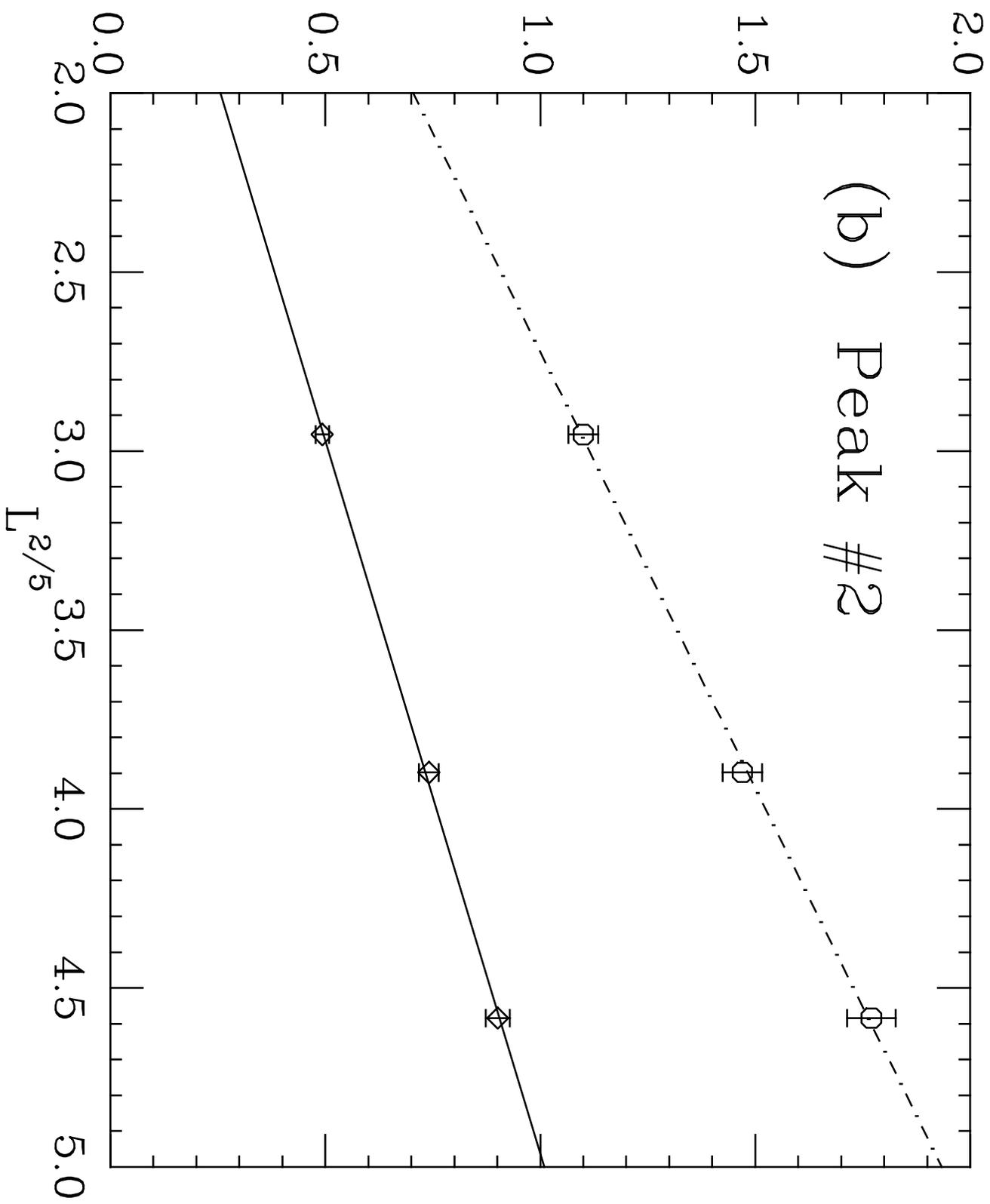
\end{document}